\documentclass[12pt]{article}
\usepackage{epsfig}

\def\bm#1{\mbox{\boldmath$#1$\unboldmath}}

\begin{document}

\begin{titlepage}

\begin{flushright}
CLNS~04/1904\\
{\tt hep-ph/0506245}\\[0.2cm]
June 23, 2005
% Revised: September 23, 2005
\end{flushright}

\vspace{0.7cm}
\begin{center}
\Large\bf\boldmath 
Advanced Predictions for Moments of the\\
$\bar B\to X_s\gamma$ Photon Spectrum
\unboldmath
\end{center}

\vspace{0.8cm}
\begin{center}
{\sc Matthias Neubert}\\
\vspace{0.7cm}
{\sl Institute for High-Energy Phenomenology\\
Newman Laboratory for Elementary-Particle Physics, Cornell University\\
Ithaca, NY 14853, U.S.A.}
\end{center}

\vspace{1.0cm}
\begin{abstract}
\vspace{0.2cm}\noindent
Based on a new, exact QCD factorization formula for the partial 
$\bar B\to X_s\gamma$ decay rate with a restriction on large photon energy, 
improved predictions are presented for the partial moments 
$\langle E_\gamma\rangle$ and 
$\langle E_\gamma^2\rangle-\langle E_\gamma\rangle^2$ of the photon spectrum 
defined with a cut $E_\gamma\ge E_0$. In the region where $\Delta=m_b-2E_0$ is 
large compared with $\Lambda_{\rm QCD}$, a theoretical description without 
recourse to shape functions can be obtained. However, for 
$\Delta\ll m_b$ it is important to separate short-distance contributions 
arising from different scales. The leading terms in the heavy-quark expansion 
of the moments receive contributions from the scales $\Delta$ and 
$\sqrt{m_b\Delta}$ only, but not from the hard scale $m_b$. For these terms, a 
complete scale separation is achieved at next-to-next-to-leading order in 
renormalization-group improved perturbation theory, including two-loop 
matching contributions and three-loop running. The results presented here can 
be used to extract the $b$-quark mass and the quantity $\mu_\pi^2$ with 
excellent theoretical precision. A fit to experimental data reported by the 
Belle Collaboration yields 
$m_b^{\rm SF}=(4.62\pm 0.10_{\rm exp}\pm 0.03_{\rm th})$\,GeV and 
$\mu_\pi^{2,{\rm SF}}=(0.11\pm 0.19_{\rm exp}\pm 0.08_{\rm th})$\,GeV$^2$ in 
the shape-function scheme at a scale $\mu_f=1.5$\,GeV, while
$m_b^{\rm kin}=(4.54\pm 0.11_{\rm exp}\pm 0.04_{\rm th})$\,GeV and 
$\mu_\pi^{2,{\rm kin}}=(0.49\pm 0.18_{\rm exp}\pm 0.09_{\rm th})$\,GeV$^2$ in 
the kinetic scheme at a scale $\mu_f=1$\,GeV.
\end{abstract}
\vfil

\end{titlepage}

\section{Introduction}

Experimental studies and theoretical analyses of inclusive decays of $B$ 
mesons have steadily been refined over the past decade. The rates for 
semileptonic $\bar B\to X\,l^-\bar\nu$ decays provide access to the elements 
$|V_{cb}|$ and $|V_{ub}|$ of the quark mixing matrix. The rate for the 
radiative decay $\bar B\to X_s\gamma$ serves as a probe for new flavor- or 
CP-violating interactions. Shape variables, such as moments of inclusive 
spectra in different kinematic variables, can be used as a tool to probe 
non-perturbative QCD dynamics in a regime where it is controllable using 
systematic heavy-quark expansions. Global fits to moments of the 
charged-lepton energy spectrum and of the invariant hadronic-mass distribution 
in $\bar B\to X_c\,l^-\bar\nu$ decay not only give the currently most precise 
determination of $|V_{cb}|$, but also of the $b$-quark mass and of other 
hadronic parameters characterizing bound-state effects in the $B$ meson, such 
as the quantity $\mu_\pi^2$ related to the $b$-quark kinetic energy 
\cite{Aubert:2004aw,Bauer:2004ve}. 

Moments of the photon energy spectrum in $\bar B\to X_s\gamma$ decay are 
another source of information about such hadronic parameters. In particular, 
while in $\bar B\to X_c\,l^-\bar\nu$ decay one is primarily sensitive to the 
quark-mass difference $(m_b-m_c)$, a measurement of the average photon energy 
in $\bar B\to X_s\gamma$ decay comes close to a direct measurement of $m_b$. 
Existing predictions for these moments rely on a conventional heavy-quark
expansion in powers of $\alpha_s(m_b)$ and $\Lambda_{\rm QCD}/m_b$. The 
usefulness of the photon-energy moments for the determination of $m_b$ and 
$\mu_\pi^2$ was first noted by Kapustin and Ligeti \cite{Kapustin:1995nr}, who 
computed the terms of order $\alpha_s$ and $1/m_b^2$ in the heavy-quark 
expansion. These authors showed that moments of the photon spectrum are in 
many aspects simpler than the spectrum itself. Perturbative corrections to the 
moments of order $\beta_0\alpha_s^2$ were calculated (in part numerically) in 
\cite{Ligeti:1999ea}, and $1/m_b^3$ corrections were studied in 
\cite{Bauer:1997fe}. Recently, an all-order resummation of the 
$\beta_0^{n-1}\alpha_s^n$ terms was performed in \cite{Benson:2004sg}. In this 
paper, the authors stress the importance of shape-function effects in the 
region where $E_0$ is larger than about 1.85\,GeV. As emphasized in 
\cite{Kagan:1998ym}, theoretical ``biases'' are introduced when moments 
measured in this region are compared with theoretical predictions obtained by 
ignoring these effects. The proposal of Benson et al.\ \cite{Benson:2004sg} is 
to correct for these biases in a {\em model-dependent way\/} by fitting a 
two-parameter shape-function model to the $\bar B\to X_s\gamma$ data, and then 
use the fitted spectrum to compute the differences between the true moments 
and the moments predicted using the conventional heavy-quark expansion 
(without shape functions). It is clear that in this way one obtains an 
accurate description of the cutoff dependence of the moments. However, the 
sensitivity of the moments to the parameters $m_b$ and $\mu_\pi^2$ now enters 
via the model ansatz used for the shape function. This introduces uncontrolled 
theoretical uncertainties, which in our opinion are underestimated in 
\cite{Benson:2004sg}. The conclusion of these authors that the naive 
heavy-quark expansion can be trusted for values $E_0<1.85$\,GeV rests on the 
model-dependent assumption that shape-function tails and other low-scale 
effects are irrelevant in that case.

Here we follow a different strategy. It is well-known that in the endpoint 
region, where $m_b-2E_\gamma\sim\Lambda_{\rm QCD}$, the $\bar B\to X_s\gamma$ 
photon spectrum obeys a QCD factorization formula of the type 
$d\Gamma/dE_\gamma\sim H\cdot J\otimes S$ 
\cite{Korchemsky:1994jb,Akhoury:1995fp}, where $H$ accounts for hard gluon 
corrections associated with the scale $m_b$, the function $J$ describes the 
properties of the final-state hadronic jet $X_s$ with invariant mass of order 
$\sqrt{m_b\Lambda_{\rm QCD}}$, and the shape function $S$ accounts for 
hadronic effects inside the $B$ meson \cite{Neubert:1993um,Bigi:1993ex}. It 
has been argued that such a formula (with functions $H_i$, $J_i$, $S_i$) holds 
not only at leading power, but at every order in the $1/m_b$ expansion 
\cite{Lee:2004ja,Bosch:2004cb,Beneke:2004in}. The question about the precise 
nature of the transition from the shape-function regime 
$m_b-2E_\gamma\sim\Lambda_{\rm QCD}$ to the kinematic region where 
$m_b-2E_\gamma\gg\Lambda_{\rm QCD}$ has recently been clarified in 
\cite{Neubert:2004dd}. A key result is that integrals over the shape function 
weighted by arbitrary smooth functions can be expanded in terms of 
local-operator matrix elements as soon as the integration domain becomes 
sufficiently large \cite{Bosch:2004th,Bauer:2003pi}. The shape function can be 
related to the discontinuity of the forward $B$-meson matrix element of the 
non-local heavy-quark effective theory (HQET) operator 
$\bar h_v (\omega+in\cdot D+i\epsilon)^{-1} h_v$, where $v$ is the $B$-meson 
velocity, and $n$ is a light-like vector satisfying $n^2=0$ and $v\cdot n=1$. 
This matrix element has a branch cut along the real axis in the complex 
$\omega$-plane, extending from $-\bar\Lambda\le\omega<\infty$, where 
$\bar\Lambda=(m_B-m_b)_{m_b\to\infty}$ is the familiar HQET parameter defined 
in the heavy-quark limit. It follows that integrals of the type
\begin{equation}\label{contour}
   \int_{-\bar\Lambda}^\Delta d\omega\,S(\omega,\mu)\,f(\omega)
   \propto \oint\limits_{|\omega|=\Delta}\!\!\!d\omega\,f(\omega)\,
   \langle\bar B(v)|\,\bar h_v \frac{1}{\omega+in\cdot D+i\epsilon}\,h_v\,
   |\bar B(v)\rangle
\end{equation}
can be written as contour integrals along a circle of radius $\Delta$ in the
complex $\omega$-plane, as long as the weight function $f(\omega)$ is analytic 
inside this circle (which is always the case in practical applications) and
$\Delta>\bar\Lambda$. In the case of the partial $\bar B\to X_s\gamma$ decay 
rate, phase space is such that the upper limit on the integral over $\omega$ 
is set by the parameter $\Delta\equiv m_b-2E_0$, where $E_0$ denotes the lower 
cut on the photon energy. For $\Delta\sim\Lambda_{\rm QCD}$ the relation above 
is not of much use. However, for $\Delta\gg\Lambda_{\rm QCD}$ the right-hand 
side of (\ref{contour}) admits an expansion in terms of $B$-meson matrix 
elements of local HQET operators. This is an expansion in powers of 
$(\Lambda_{\rm QCD}/\Delta)^n$, i.e., {\rm not\/} a conventional heavy-quark 
expansion. The corresponding Wilson coefficients have a perturbative expansion 
in powers of $\alpha_s(\mu)$, which is free of large logarithms provided that 
the renormalization scale used in the definition of the renormalized shape 
function is chosen such that $\mu\sim\Delta$. The scale dependence of the 
leading-order shape function $S(\omega,\mu)$ can be controlled precisely, 
because an exact analytic solution to its evolution equation exists in 
momentum space \cite{Neubert:2004dd,Bosch:2004th,Lange:2003ff}.

In previous work, we have applied this technology to derive a 
renormalization-group (RG) improved prediction for the partial 
$\bar B\to X_s\gamma$ decay rate as a function of the photon cut for values of 
$E_0$ outside the shape-function region (typically $E_0<2$\,GeV, such that 
$\Delta>0.7$\,GeV) \cite{Neubert:2004dd}. In that paper we have already 
presented a formula for the first moment of the photon spectrum. An important
finding was that the first two terms in the $1/m_b$ expansion of the average
photon energy do not receive any contributions from the hard scale $m_b$. Here 
we extend this analysis to the second moment. Most importantly, we include the 
complete set of two-loop matching corrections and three-loop 
anomalous-dimension effects so as to obtain predictions that are exact at 
next-to-next-to-leading order (NNLO) in RG-improved perturbation theory. We 
confirm that, in general, moments of the photon spectrum probe low-scale 
dynamics sensitive to the scales $\mu_0\sim\Delta$ and 
$\mu_i\sim\sqrt{m_b\Delta}$. To a very good approximation, they are 
insensitive to physics at the scale $m_b$. As long as $\Delta\ll m_b$ 
($\Delta\approx 1$\,GeV for present experiments), it is thus not appropriate 
to compute the moments using a conventional heavy-quark expansion in powers of 
$\alpha_s(m_b)$ and $\Lambda_{\rm QCD}/m_b$. Compared with 
\cite{Neubert:2004dd} we also include additional small corrections arising at
higher orders in the $1/m_b$ expansion, and we comment on the effects of the 
photon-energy cut on the Lorentz boost between the $B$-meson rest frame and 
the $\Upsilon(4S)$ rest frame, which must be corrected for in the experimental 
analyses \cite{Chen:2001fj,Koppenburg:2004fz,Aubert:2005cu}.

\section{Factorization formula for the decay rate}

\subsection{Partial decay rate and moment relations}
\label{sec:sec2intro}

We begin by collecting some useful relations for moments of the partial 
$\bar B\to X_s\gamma$ decay rate, defined with a cut $E_\gamma\ge E_0$ on the 
photon energy measured in the $B$-meson rest frame. It is convenient to define 
a variable $p_+=m_b-2E_\gamma$, where for the time being $m_b$ denotes the 
pole mass of the heavy $b$ quark. The requirement $E_\gamma\ge E_0$ translates 
into $p_+\le\Delta=m_b-2E_0$. As long as $\Delta$ is not too small, the 
partial rate can be calculated using an operator product expansion (OPE). From 
(\ref{contour}) it follows that the correct criterion for the validity of the 
OPE is $\Delta\gg\Lambda_{\rm QCD}$. We define
\begin{equation}\label{eq2}
   \Gamma_{\rm OPE}(\Delta)
   = \int_0^\Delta\!dp_+\,\frac{d\Gamma_{\rm OPE}}{dp_+} \,,
\end{equation} 
taking into account that in the OPE the phase space is such that 
$0\le p_+\le m_b$. It follows from this expression that partial moments in the 
variable $p_+$, defined as
\begin{equation}
   \langle p_+^n\rangle =
   \frac{\displaystyle
         \int_0^\Delta\!dp_+\,p_+^n\,\frac{d\Gamma_{\rm OPE}}{dp_+}}
        {\displaystyle
         \int_0^\Delta\!dp_+\,\frac{d\Gamma_{\rm OPE}}{dp_+}} 
   \equiv \Delta^n\,M_n(\Delta) \,,
\end{equation}
can be written in terms of integrals over the function 
$\Gamma_{\rm OPE}(\Delta)$, namely
\begin{equation}\label{Mnformula}
   M_n(\Delta) = 1 - \frac{n}{\Gamma_{\rm OPE}(\Delta)}
   \int_0^1\!dy\,y^{n-1}\,\Gamma_{\rm OPE}(y\Delta) \,.
\end{equation}
Given a theoretical formula for the partial rate $\Gamma_{\rm OPE}(\Delta)$, 
it is thus possible to derive arbitrary moments without going back to the 
differential spectrum itself. Note that for the application of this relation 
it is irrelevant that the variable $y\Delta$ is not always large compared with 
$\Lambda_{\rm QCD}$. As in (\ref{eq2}), what matters is only the upper limit 
of the integration domain, because this sets the radius of the corresponding
contour integral.

Moments of the photon energy spectrum can immediately be related to the 
functions $M_n(\Delta)$. In particular, central moments defined with respect 
to the cutoff, i.e.\ $\langle(E_\gamma-E_0)^n\rangle$, are linear combinations 
of the $M_n(\Delta)$. For the average photon energy and the variance 
$\sigma_E^2\equiv\langle E_\gamma^2\rangle-\langle E_\gamma\rangle^2$ of the 
photon spectrum, we then obtain
\begin{eqnarray}\label{physmoments}
   \langle E_\gamma \rangle - E_0
   &=& \frac{\Delta}{2} \left[ 1 - M_1(\Delta) \right] , \nonumber\\
   \sigma_E^2
   &=& \frac{\Delta^2}{4} \left[ 1 - 2 M_1(\Delta) + M_2(\Delta) \right]
    - \left( \langle E_\gamma \rangle - E_0 \right)^2 .
\end{eqnarray}
The main goal of this work is to derive accurate theoretical expressions for 
these moments in a region where the cutoff $E_0$ is such that 
$\Lambda_{\rm QCD}\ll\Delta\ll m_b$. The first inequality 
($\Lambda_{\rm QCD}\ll\Delta$) ensures that theoretical predictions can be 
obtained without recourse to non-perturbative shape (or bias) functions. 
The second inequality ($\Delta\ll m_b$) implies that the theory used to derive 
these predictions cannot be a conventional heavy-quark expansion in powers of 
$\alpha_s(m_b)$ and $\Lambda_{\rm QCD}/m_b$. Instead, one should disentangle 
the physics associated with the different short-distance scales 
$\Delta\ll\sqrt{m_b\Delta}\ll m_b$. For the partial rate 
$\Gamma_{\rm OPE}(\Delta)$, this has been achieved (at leading power in 
$1/m_b$) in \cite{Neubert:2004dd}.

Already at this stage it is instructive to ask what precision we might expect 
to achieve in the calculation of $\langle E_\gamma\rangle$ and $\sigma_E^2$. 
We will see below that the moments $M_n(\Delta)$ have an expansion in powers 
of $\alpha_s$, $\Lambda_{\rm QCD}/\Delta$, and $\Delta/m_b$. Since the leading 
terms in the expansion are known at two-loop order, it is reasonable to expect 
a precision on $M_n(\Delta)$ of about 3\%. With $\Delta\approx 1$\,GeV, it 
follows that $\delta\langle E_\gamma\rangle\approx 0.015$\,GeV and 
$\delta\sigma_E^2\approx 0.08$\,GeV$^2$. At tree level, the theoretical 
expressions for the moments are $\langle E_\gamma\rangle=m_b/2+\dots$ and 
$\sigma_E^2=\mu_\pi^2/12+\dots$, so that we may expect to extract the 
heavy-quark parameters with precision $\delta m_b\approx 30$\,MeV and 
$\delta\mu_\pi^2\approx 0.1$\,GeV$^2$. These estimates will be confirmed by 
the more elaborate study in Section~\ref{sec:numerics}. While the projected 
accuracy for the $b$-quark mass determination is exquisite, the extraction of 
$\mu_\pi^2$ suffers to some extent from the fact that the hadronic 
contribution $\mu_\pi^2/12$ to the variance $\sigma_E^2$ competes with a large 
perturbative ``background'' of order $\alpha_s\Delta^2$.

\subsection{A wonderful formula}

At leading power in $1/m_b$ and next-to-leading order (NLO) in the expansion 
in powers of $\Lambda_{\rm QCD}/\Delta$, it is possible to write an expression 
for the partial rate $\Gamma_{\rm OPE}(\Delta)$ that is valid to all orders in
perturbation theory, and in which nevertheless the dependence on the variable
$\Delta$ enters in a very transparent way. Starting point is the factorization 
formula \cite{Neubert:2004dd}\footnote{The variable $\omega$ corresponds to 
$\hat\omega-\bar\Lambda$ in the notation of \cite{Neubert:2004dd}, and to 
$-\omega$ in the notation of \cite{Bosch:2004th}.}
\begin{eqnarray}\label{ff}
   \Gamma(\Delta)
   &=& \frac{G_F^2\alpha}{32\pi^4}\,|V_{tb} V_{ts}^*|^2\,m_b^3\,
    \overline{m}_b^2(\mu_h)\,|H_\gamma(\mu_h)|^2\,U_1(\mu_h,\mu_i)\,
    U_2(\mu_i,\mu_0)\,\frac{e^{-\gamma_E\eta}}{\Gamma(1+\eta)} \nonumber\\
   &\times& \eta \int_{-\bar\Lambda}^{\Delta} dp_+
    \int_{-\bar\Lambda}^{p_+} d\omega\,
    m_b\,J(m_b(p_+ -\omega),\mu_i)
    \int_{-\bar\Lambda}^{\omega} d\omega'\,
    \frac{S(\omega',\mu_0)}{\mu_0^\eta(\omega-\omega')^{1-\eta}} 
    + \dots \,,
\end{eqnarray}
where the ellipses represent power corrections in $1/m_b$. Here $m_b$ is the 
$b$-quark pole mass, and $\overline{m}_b(\mu)$ denotes the running mass 
defined in the $\overline{\rm MS}$ scheme. The function $H_\gamma$ contains 
hard quantum fluctuations associated with the weak-interaction vertices in the 
effective weak Hamiltonian. The jet function $J$ describes the physics of the 
hadronic final state $X_s$. The shape function $S$ governs the soft physics 
associated with bound-state effects inside the $B$ meson. The matching scales 
$\mu_h\sim m_b$, $\mu_i\sim\sqrt{m_b\Delta}$, and $\mu_0\sim\Delta$ serve to 
separate the hard, hard-collinear, and soft components in the factorization 
theorem, and the RG functions $U_1$ and $U_2$ resum logarithmic corrections 
arising from evolution between these scales. The precise form of these 
objects, which can be found in \cite{Neubert:2004dd}, is irrelevant to our 
discussion. Finally, the variable
\begin{equation}\label{eta}
   \eta = 2\int_{\mu_0}^{\mu_i} \frac{d\mu}{\mu}\,
   \Gamma_{\rm cusp}[\alpha_s(\mu)]
\end{equation}
is given in terms of an integral over the universal cusp anomalous dimension 
of Wilson loops with light-like segments \cite{Korchemsky:1987wg}. The 
perturbative expansion of this quantity is discussed in Appendix~A.1. The 
result (\ref{ff}) is formally independent of the choices of the matching 
scales. In practice, a residual scale dependence remains because one is forced 
to truncate the perturbative expansions of the various objects in the 
factorization formula.

Introducing the integral 
\begin{equation}\label{jdef}
   j\bigg( \ln\frac{Q^2}{\mu^2},\mu \bigg)\equiv \int_0^{Q^2}\!dk^2\,J(k^2,\mu)
\end{equation}
over the jet function, and changing the order of the integrations over $p_+$ 
and $\omega$, the terms in the second line of (\ref{ff}) can be rewritten in 
the form
\begin{equation}
   {\bf j}\bigg( \ln\frac{m_b\mu_0}{\mu_i^2} + \partial_\eta,\mu_i\bigg)
   \int_{-\bar\Lambda}^{\Delta} d\omega\,S(\omega,\mu_0)
   \left( \frac{\Delta-\omega}{\mu_0} \right)^\eta ,
\end{equation}
where we have defined
\begin{equation}\label{jhatdef}
   {\bf j}(L,\mu)\equiv \eta\int_0^1 \frac{dz}{z^{1-\eta}}\,
   j\Big( L + \ln(1-z),\mu \Big) \,.
\end{equation}
When $L$ contains a derivative operator $\partial_\eta$, it is understood that 
this operator acts only to the right. 

The remaining shape-function integral is of the type shown in (\ref{contour}), 
and for $\Delta\gg\Lambda_{\rm QCD}$ it can be expanded in matrix elements of 
local HQET operators. To this end, we replace the true shape function by the
corresponding function obtained in the parton model with on-shell external 
quark states, which has support for $\omega\ge -n\cdot k$ instead of 
$\omega\ge-\bar\Lambda$, where $k=p_b-m_b v$ with $v\cdot k=0$ is the residual 
momentum of the on-shell heavy quark. We then evaluate the integral, expand 
the result in powers of $n\cdot k$, and match the answer onto HQET matrix 
elements \cite{Bosch:2004th}. It is convenient to use an integration by parts 
and introduce the integral
\begin{equation}\label{sdef}
   s\bigg( \ln\frac{\Omega_k}{\mu},\mu \bigg)
   \equiv \int_{-n\cdot k}^{\Omega} d\omega\,S_{\rm parton}(\omega,\mu) \,,
\end{equation}
which is analogous to the function $j$ in (\ref{jdef}). Reparameterization 
invariance \cite{Luke:1992cs,Falk:1992fm} ensures that the result only depends 
on the sum $\Omega_k=\Omega+n\cdot k$. Introducing a related function
\begin{equation}
   {\bf s}(L,\mu)\equiv \eta\int_0^1 \frac{dz}{z^{1-\eta}}\,
   s\Big( L + \ln(1-z),\mu \Big)
\end{equation}
in analogy with (\ref{jhatdef}), we find that
\begin{eqnarray}
   \int_{-\bar\Lambda}^{\Delta} d\omega\,S_{\rm parton}(\omega,\mu_0)
   \left( \frac{\Delta-\omega}{\mu_0} \right)^\eta 
   &=& \frac{\eta}{\mu_0} \int_{-n\cdot k}^{\Delta}\!d\omega
    \left( \frac{\Delta-\omega}{\mu_0} \right)^{\eta-1}
    s\bigg( \ln\frac{\omega+n\cdot k}{\mu_0},\mu_0 \bigg) \nonumber\\
   &=& {\bf s}\left(\partial_\eta,\mu_0\right)
    \left( \frac{\Delta+n\cdot k}{\mu_0} \right)^\eta .
\end{eqnarray}
We now expand this result to second order in $n\cdot k$ and replace 
$n\cdot k\to 0$, $(n\cdot k)^2\to\mu_\pi^2/3$, which accomplishes the matching
onto local HQET matrix elements to first non-trivial order. This yields to the 
following result for the terms in the second line of the factorization 
formula~(\ref{ff}):
\begin{equation}
   {\bf j}\bigg( \ln\frac{m_b\mu_0}{\mu_i^2} + \partial_\eta,\mu_i\bigg)\,
   {\bf s}\left(\partial_\eta,\mu_0\right)
   \left( \frac{\Delta}{\mu_0} \right)^\eta \left[
   1 - \frac{\eta(1-\eta)}{6}\,\frac{\mu_\pi^2}{\Delta^2} + \dots \right] .
\end{equation}

It remains to derive the explicit form of the functions ${\bf j}$ and 
${\bf s}$. This can be accomplished by noting that at any order in 
perturbation theory the objects $j(L,\mu)$ and $s(L,\mu)$ are polynomials in 
$L$ (see Section~\ref{sec:2loopcis} below), so that it suffices to compute the 
integrals
\begin{eqnarray}
   &&\eta \int_0^1 \frac{dz}{z^{1-\eta}}
    \left( L + \partial_\eta + \ln(1-z) \right)^n 
   = \partial_\epsilon^n \int_0^1\!dz\,\eta\,z^{\eta-1}
    \left( (1-z)\,e^{L + \partial_\eta} \right)^\epsilon \bigg|_{\epsilon=0}
    \nonumber\\
   &=& \partial_\epsilon^n\,
    \frac{\Gamma(1+\eta)\,\Gamma(1+\epsilon)}{\Gamma(1+\eta+\epsilon)}\,
    e^{\epsilon(L+\partial_\eta)} \bigg|_{\epsilon=0}
   \equiv \frac{\Gamma(1+\eta)}{e^{-\gamma_E\eta}}\,
    I_n(L+\partial_\eta)\,\frac{e^{-\gamma_E\eta}}{\Gamma(1+\eta)} \,,
\end{eqnarray}
where
\begin{equation}
   I_n(x) = \partial_\epsilon^n
    \left[ e^{\epsilon(x+\gamma_E)}\,\Gamma(1+\epsilon) \right]_{\epsilon=0}
   = \partial_\epsilon^n \exp\left[ \epsilon x + \sum_{k=2}^\infty\,
   \frac{(-1)^k}{k}\,\epsilon^k \zeta_k \right]_{\epsilon=0}
\end{equation}
are polynomials of degree $n$. For our purposes we need the first four of 
them, which are
\begin{eqnarray}
    I_1(x) &=& x \,, \hspace{2.0cm}
     I_3(x) = x^3 + \frac{\pi^2}{2}\,x - 2\zeta_3 \,, \nonumber\\
    I_2(x) &=& x^2 + \frac{\pi^2}{6} \,, \qquad
     I_4(x) = x^4 + \pi^2 x - 8\zeta_3\,x + \frac{3\pi^4}{20} \,.
\end{eqnarray}
We now define functions $\widetilde s$ and $\widetilde j$ by the 
following replacement rules:
\begin{equation}\label{widetildejs}
   \widetilde j(L,\mu)\equiv j(L,\mu) \Big|_{L^n\to I_n(L)} \,, \qquad
   \widetilde s(L,\mu)\equiv s(L,\mu) \Big|_{L^n\to I_n(L)} \,.
\end{equation}
It follows that 
\begin{equation}
   {\bf j}(L+\partial_\eta,\mu) = \frac{\Gamma(1+\eta)}{e^{-\gamma_E\eta}}\,
   \widetilde j(L+\partial_\eta,\mu)\,
   \frac{e^{-\gamma_E\eta}}{\Gamma(1+\eta)} \,,
\end{equation}
and similarly for the soft function. The exact result for the leading-power 
contribution to the partial decay rate in the OPE can now be written in the 
remarkable form
\begin{eqnarray}\label{GOPE}
   &&\hspace{-0.6cm}\Gamma_{\rm OPE}(\Delta)
    = \frac{G_F^2\alpha}{32\pi^4}\,|V_{tb} V_{ts}^*|^2\,m_b^3\,
    \overline{m}_b^2(\mu_h)\,|H_\gamma(\mu_h)|^2\,
    U_1(\mu_h,\mu_i)\,U_2(\mu_i,\mu_0) \\
   &\times& \left\{ \widetilde j\bigg( \ln\frac{m_b\mu_0}{\mu_i^2}
    + \partial_\eta,\mu_i \bigg)\,
    \widetilde s\left(\partial_\eta,\mu_0\right) 
    \frac{e^{-\gamma_E\eta}}{\Gamma(1+\eta)}
    \left( \frac{\Delta}{\mu_0} \right)^\eta \left[
    1 - \frac{\eta(1-\eta)}{6}\,\frac{\mu_\pi^2}{\Delta^2} + \dots \right]
    + {\cal O}\left( \frac{\Delta}{m_b} \right) \right\} . \nonumber
\end{eqnarray}
This result implies that for $m_b-2E_\gamma\gg\Lambda_{\rm QCD}$ the photon 
spectrum exhibits a radiation tail, 
$d\Gamma/dE_\gamma\propto 1/(m_b-2E_\gamma)^{1-\eta}$ modulo small logarithmic 
corrections, which falls off slowly with energy. The presence of this tail and 
its phenomenological implications have been discussed in 
\cite{Neubert:2004dd}. Note that even at leading power in $1/m_b$ there exist 
non-perturbative corrections of the form $(\Lambda_{\rm QCD}/\Delta)^n$ with 
$n\ge 2$. We have included the leading such term, which is proportional to the 
kinetic-energy parameter $\mu_\pi^2$. For $\Delta\sim 1$\,GeV the numerical 
impact of these power corrections is rather small, so it seems safe to 
truncate the series at this point.  

It is worth emphasizing that expression (\ref{GOPE}) is exact to all orders in 
perturbation theory, and it is valid for any values of the matching scales 
$\mu_h$, $\mu_i$, and $\mu_0$. In the limit where all three matching scale are 
set equal to a common scale $\mu$, the result smoothly reduces to conventional 
fixed-order perturbation theory. While the resummed, factorized expression is 
superior to a fixed-order result whenever there are widely separated scales in 
the problem (i.e., for $\Delta\ll m_b$), it remains valid in the limit where 
the different scales become of the same order ($\Delta\sim m_b$).\footnote{In 
this case, of course, power corrections of order $\Delta/m_b$ would become 
important.} 
In other words, there is never a region where fixed-order perturbation theory 
would be more appropriate to use than the above factorization formula.

\subsection{Evolution equations and two-loop results}
\label{sec:2loopcis}

The functions $j(L,\mu)$ and $s(L,\mu)$ obey integro-differential RG 
equations, which can be derived starting from the evolution equations for the
jet and shape functions discussed in \cite{Neubert:2004dd,Bosch:2004th}. We 
find 
\begin{eqnarray}\label{RGE}
   \frac{d}{d\ln\mu}\,j(L,\mu)
   &=& - 2 \Big[ \Gamma_{\rm cusp}(\alpha_s)\,L + \gamma^J(\alpha_s)
    \Big]\,j(L,\mu) \nonumber\\
   &&\mbox{}- 2 \Gamma_{\rm cusp}(\alpha_s)
    \int_0^1 \frac{dz}{z} \Big[ j(L+\ln(1-z),\mu) - j(L,\mu) \Big] \,, 
    \nonumber\\
   \frac{d}{d\ln\mu}\,s(L,\mu)
   &=& 2 \Big[ \Gamma_{\rm cusp}(\alpha_s)\,L - \gamma(\alpha_s) \Big]\,
    s(L,\mu) \nonumber\\
   &&\mbox{}+ 2 \Gamma_{\rm cusp}(\alpha_s)
    \int_0^1 \frac{dz}{z} \Big[ s(L+\ln(1-z),\mu) - s(L,\mu) \Big] \,,
\end{eqnarray}
where $\alpha_s\equiv\alpha_s(\mu)$. We encounter again the cusp anomalous 
dimension $\Gamma_{\rm cusp}$, and in addition anomalous dimensions $\gamma$ 
and $\gamma^J$ governing the single-logarithmic evolution of the shape and jet 
functions, respectively. (Recall that for $j$ we have $L=\ln(Q^2/\mu^2)$, 
while for $s$ we have instead $L=\ln(\Omega_k/\mu)$.) These equations can be 
solved order by order in perturbation theory with a double-logarithmic 
expansion of the form
\begin{eqnarray}
   j(L,\mu) &=& 1 + \sum_{n=1}^\infty
    \left( \frac{\alpha_s(\mu)}{4\pi} \right)^n
    \left( b_0^{(n)} + b_1^{(n)} L + \dots + b_{2n-1}^{(n)} L^{2n-1}
    + b_{2n}^{(n)} L^{2n} \right) , \nonumber\\
   s(L,\mu) &=& 1 + \sum_{n=1}^\infty
    \left( \frac{\alpha_s(\mu)}{4\pi} \right)^n
    \left( c_0^{(n)} + c_1^{(n)} L + \dots + c_{2n-1}^{(n)} L^{2n-1}
    + c_{2n}^{(n)} L^{2n} \right) .
\end{eqnarray}
The evolution equations (\ref{RGE}) allow us to express all coefficients of 
logarithms in terms of the perturbative expansion coefficients of the 
anomalous dimensions and $\beta$ function. At two-loop order, we obtain
\begin{eqnarray}\label{ansatz}
   j(L,\mu) &=& 1 + \frac{\alpha_s(\mu)}{4\pi} \left[
    b_0^{(1)} + \gamma_0^J L + \frac12\,\Gamma_0 L^2 \right] \nonumber\\
   &&\mbox{}+ \left( \frac{\alpha_s(\mu)}{4\pi} \right)^2 \Bigg[
    b_0^{(2)} + \left( b_0^{(1)} (\gamma_0^J-\beta_0) + \gamma_1^J
    - \frac{\pi^2}{6}\,\Gamma_0\gamma_0^J + \zeta_3\Gamma_0^2
    \right) L \nonumber\\
   &&\mbox{}+ \frac12 \left( \gamma_0^J (\gamma_0^J-\beta_0)
    + b_0^{(1)}\,\Gamma_0 + \Gamma_1
    - \frac{\pi^2}{6}\,\Gamma_0^2 \right) L^2
    - \left( \frac16\,\beta_0 - \frac12\gamma_0^J \right)
    \Gamma_0 L^3 + \frac18\,\Gamma_0^2 L^4 \Bigg] \,, \nonumber\\
   s(L,\mu) &=& 1 + \frac{\alpha_s(\mu)}{4\pi} \left[
    c_0^{(1)} + 2\gamma_0 L - \Gamma_0 L^2 \right] \nonumber\\
   &&\mbox{}+ \left( \frac{\alpha_s(\mu)}{4\pi} \right)^2 \Bigg[
    c_0^{(2)} + \left( 2c_0^{(1)} (\gamma_0-\beta_0) + 2\gamma_1
    + \frac{2\pi^2}{3}\,\Gamma_0\gamma_0 + 4\zeta_3\Gamma_0^2 \right) L \\
   &&\mbox{}+ \left( 2\gamma_0(\gamma_0-\beta_0) - c_0^{(1)}\Gamma_0
    - \Gamma_1 - \frac{\pi^2}{3}\,\Gamma_0^2 \right) L^2 
    + \left( \frac23\,\beta_0 - 2\gamma_0 \right) \Gamma_0 L^3
    + \frac12\,\Gamma_0^2 L^4 \Bigg] \,, \nonumber
\end{eqnarray}
where \cite{Bosch:2004th,Bauer:2003pi}
\begin{equation}\label{b0c0}
   b_0^{(1)} = (7-\pi^2)\,C_F \,, \qquad
   c_0^{(1)} = - \frac{\pi^2}{6}\,C_F \,,
\end{equation}
and the coefficients $b_0^{(2)}$ and $c_0^{(2)}$ are unknown. The relevant 
expansion coefficients of the anomalous dimensions and $\beta$ functions are 
listed in Appendices~A.1 and A.2.

\section{Ingredients of the moment calculation}
\label{sec:moments}

\subsection{Results at leading power in \boldmath$1/m_b$\unboldmath}

While the result (\ref{GOPE}) is of considerable complexity when expanded 
beyond the leading order in RG-improved perturbation theory, it is well suited 
for computing the moments $M_n(\Delta)$ using relation (\ref{Mnformula}). The 
reason is that, {\em before\/} the derivatives with respect to $\eta$ are 
carried out, the dependence on $\Delta$ is of power type. According to 
(\ref{Mnformula}) the moments $M_n(\Delta)$ are obtained from ratios of 
expressions linear in the decay rate, and hence any $\Delta$-independent 
factors cancel out. It follows that the entire first line in (\ref{GOPE}), and 
in particular all reference to the hard scale $\mu_h$, cancels in the formulae 
for the moments. (A very weak dependence on the hard scale enters through the 
$1/m_b$ corrections, see below.) Also, after the integral over $y$ in 
(\ref{Mnformula}) has been carried out, the product 
$[e^{-\gamma_E\eta}/\Gamma(1+\eta)]\,(\Delta/\mu_0)^\eta$ can be pulled 
through the differential operators $\widetilde s$ and $\widetilde j$ using 
that
\begin{equation}
   \partial_\eta\,\frac{e^{-\gamma_E\eta}}{\Gamma(1+\eta)}
   \left( \frac{\Delta}{\mu_0} \right)^\eta f(\eta)
   = \frac{e^{-\gamma_E\eta}}{\Gamma(1+\eta)}
   \left( \frac{\Delta}{\mu_0} \right)^\eta 
   \left( \ln\frac{\Delta}{\mu_0} - h(\eta) + \partial_\eta \right) f(\eta)
   \,, 
\end{equation}
where $h(\eta)=\psi(1+\eta)+\gamma_E$. It is then immediate to obtain the 
following, exact form for the moments at leading power in $1/m_b$, indicated 
with a superscript ``(0)'':
\begin{equation}\label{Mnmaster}
   M_n^{(0)}(\Delta) = 1 - 
   \frac{\displaystyle {\cal D}(\nabla_\eta) 
         \left[ \frac{n}{n+\eta} - \frac{n}{n+\eta-2}\,
         \frac{\eta(1-\eta)}{6}\,\frac{\mu_\pi^2}{\Delta^2} + \dots \right]}
        {\displaystyle {\cal D}(\nabla_\eta)
         \left[ 1 - \frac{\eta(1-\eta)}{6}\,\frac{\mu_\pi^2}{\Delta^2} + \dots
         \right]} \,.
\end{equation}
The object
\begin{equation}
   {\cal D}(\nabla_\eta)
   = \widetilde j\bigg( \ln\frac{m_b\Delta}{\mu_i^2} + \nabla_\eta \bigg)\,
   \widetilde s\big( \ln\frac{\Delta}{\mu_0} + \nabla_\eta \big) \,; \qquad
   \nabla_\eta = \frac{d}{d\eta} - h(\eta)
\end{equation}
is a differential operator defined in terms of the functions $\widetilde j$ 
and $\widetilde s$, which are determined in terms of the matching corrections 
at the hard-collinear and soft scales, $\mu_i$ and $\mu_0$, respectively. A 
careful analysis of the equations that led to (\ref{GOPE}) shows that the 
result for the $\mu_\pi^2$ term in the numerator of (\ref{Mnmaster}) is 
correct for any positive integer $n$, even though the integral over $y$ in 
(\ref{Mnformula}) appears to diverge for $n<2$. At two-loop order 
${\cal D}(\nabla_\eta)$ is a fourth-order polynomial in $\nabla_\eta$. It is 
understood that the derivatives with respect to $\eta$ in an expression of the 
form ${\cal D}(\nabla_\eta)\,f(\eta)$ act on both $f(\eta)$ and the function 
$h(\eta)$ in the definition of $\nabla_\eta$, e.g.\ $\nabla_\eta^2\,f(\eta)%
=f''(\eta)-2h(\eta)\,f'(\eta)-h'(\eta)\,f(\eta)+h^2(\eta)\,f(\eta)$. Note
the important fact that the unknown two-loop coefficients $b_0^{(2)}$ and
$c_0^{(2)}$ in (\ref{ansatz}) cancel in the ratio (\ref{Mnmaster}). This means 
that we have all the ingredients in place to obtain predictions for the 
moments $M_n$ that are valid at NNLO in RG-improved perturbation theory. At 
this order, exact two-loop matching conditions at the scales $\mu_i$ and 
$\mu_0$ are combined with three-loop running effects incorporated in the 
calculation of the parameter $\eta$, which resums logarithms of the ratio 
$(\mu_i/\mu_0)^2\sim m_b/\Delta$. 

Eqs.~(\ref{GOPE}) and (\ref{Mnmaster}) are the main results of this work. It 
is straightforward to work out the leading-power contributions to the moments 
$M_1(\Delta)$ and $M_2(\Delta)$ by carrying out the derivatives with respect 
to $\eta$ in (\ref{Mnmaster}) and expanding the resulting expression 
consistently in powers of $\alpha_s(\mu_i)$, $\alpha_s(\mu_0)$ and to first 
order in $\mu_\pi^2/\Delta^2$. At NLO in RG-improved perturbation theory, we 
find
\begin{eqnarray}\label{M2lo}
   M_1^{(0)}(\Delta)
   &=& \frac{1}{(1+\eta)^2}\,\Bigg\{ 
    \left( 1 - \frac{\mu_\pi^2}{3\Delta^2} \right)
    \Bigg[ \eta(1+\eta) + \frac{C_F\alpha_s(\mu_i)}{\pi} \left(
    \ln\frac{m_b\Delta}{\mu_i^2} - h(\eta) - \frac34 - \frac{1}{1+\eta}
    \right) \nonumber\\
   &&\hspace{1.7cm}\mbox{}+ \frac{C_F\alpha_s(\mu_0)}{\pi} \left(
    - 2\ln\frac{\Delta}{\mu_0} + 2h(\eta) - 1 + \frac{2}{1+\eta} \right)
    \Bigg] \nonumber\\
   &&\hspace{1.7cm}\mbox{}+ \frac{\mu_\pi^2}{3\Delta^2}\,(1-2\eta)
    \left[ \frac{C_F\alpha_s(\mu_0)}{\pi} - \frac{C_F\alpha_s(\mu_i)}{2\pi}
    \right] \Bigg\} \,, \nonumber\\
   M_2^{(0)}(\Delta)
   &=& \frac{2}{(2+\eta)^2}\,\Bigg\{ 
    \left( 1 - \frac{\mu_\pi^2}{\Delta^2} \right)
    \Bigg[ \frac{\eta(2+\eta)}{2} + \frac{C_F\alpha_s(\mu_i)}{\pi} \left(
    \ln\frac{m_b\Delta}{\mu_i^2} - h(\eta) - \frac34 - \frac{1}{2+\eta}
    \right) \nonumber\\
   &&\hspace{1.7cm}\mbox{}+ \frac{C_F\alpha_s(\mu_0)}{\pi} \left(
    - 2\ln\frac{\Delta}{\mu_0} + 2h(\eta) - 1 + \frac{2}{2+\eta} \right)
    \Bigg] \nonumber\\
   &&\hspace{1.7cm}\mbox{}+ \frac{\mu_\pi^2}{3\Delta^2}\,(1-2\eta)
    \left[ \frac{C_F\alpha_s(\mu_0)}{\pi} - \frac{C_F\alpha_s(\mu_i)}{2\pi}
    \right] \Bigg\} 
    + \frac{\mu_\pi^2}{3\Delta^2} \,.
\end{eqnarray}
The corresponding expressions valid at NNLO are far more complicated. We do 
not list them explicitly, as it is easiest to generate them directly from 
(\ref{Mnmaster}). In the resulting formulae one should expand the quantity 
$\eta$ consistently to the required order in RG-improved perturbation theory, 
using the results compiled in Appendix~A.1. (Using instead the NNLO expression 
for $\eta$ everywhere makes a negligible difference in our numerical results.)
Also, before applying these results to the analysis of experimental data, the 
pole-scheme parameters $m_b$ and $\mu_\pi^2$ should be eliminated in favor of 
corresponding parameters defined in a physical renormalization scheme.

The scale separation achieved using RG techniques, which allows us to 
disentangle the physics at the hard, hard-collinear, and soft scales ($m_b$, 
$\sqrt{m_b\Delta}$, and $\Delta$), is one of the most important ingredients of 
our approach. This, combined with the fact that for the first time we include 
the complete two-loop perturbative corrections, distinguishes our calculation 
from all previous analyses of the $\bar B\to X_s\gamma$ moments. The physical 
insight that the shape variables probe low-scale dynamics, while at leading 
power in $1/m_b$ they are insensitive to physics at the hard scale $m_b$, 
makes it apparent that a precise control over low-scale perturbative 
corrections is crucial to obtain reliable predictions for the moments. 

In order to compare our RG-improved results with those of the conventional 
heavy-quark approach, it is useful to expand expression (\ref{Mnmaster}) to 
two-loop order in fixed-order perturbation theory. We obtain
\begin{eqnarray}
   M_1^{(0)}(\Delta)
   &=& \left( 1 - \frac{\mu_\pi^2}{3\Delta^2} \right)
    \frac{C_F\alpha_s(\mu)}{\pi} \left[ - \ln\frac{\Delta}{m_b} - \frac34 
    + \frac{\alpha_s(\mu)}{\pi}\,k_1(\Delta) \right] \nonumber\\
   &&\mbox{}+ \frac{\mu_\pi^2}{3\Delta^2}\,\frac{C_F\alpha_s(\mu)}{\pi} 
    \left[\, \frac12 + \frac{\alpha_s(\mu)}{\pi}\,k_2(\Delta) \right] ,
    \nonumber\\
   M_2^{(0)}(\Delta)
   &=& \left( 1 - \frac{\mu_\pi^2}{\Delta^2} \right)
    \frac{C_F\alpha_s(\mu)}{\pi} \left[ - \frac12 \ln\frac{\Delta}{m_b}
    - \frac58 + \frac{\alpha_s(\mu)}{\pi}\,k_3(\Delta) \right] \nonumber\\
   &&\mbox{}+ \frac{\mu_\pi^2}{3\Delta^2}\,\Bigg[\, 1
    + \frac{C_F\alpha_s(\mu)}{4\pi}
    + C_F \left( \frac{\alpha_s(\mu)}{\pi} \right)^2 k_4(\Delta) \Bigg] \,,
\end{eqnarray}
where the two-loop coefficients are given by
{\small
\begin{eqnarray}
   k_1(\Delta)
   &=& \left( \frac{3\beta_0}{8} - C_F \right) \ln^2\frac{\Delta}{m_b}
    + \left[
    \left( \frac12 \ln\frac{m_b}{\mu} - \frac{23}{48} \right) \beta_0 
    - \left( \frac12 + \frac{\pi^2}{6} \right) C_F 
    - \left( \frac13 - \frac{\pi^2}{12} \right) C_A \right]
    \ln\frac{\Delta}{m_b} \nonumber\\
   &&\mbox{}+ \left( \frac38 \ln\frac{m_b}{\mu} - \frac{13}{32}
    + \frac{\pi^2}{24} \right) \beta_0 
    - \left( \frac{29}{32} + \frac{\zeta_3}{2} \right) C_F
    + \left( \frac{7}{16} + \frac{\zeta_3}{4} \right) C_A \,, \nonumber\\
   k_2(\Delta)
   &=& \left( 2 C_F - \frac{3\beta_0}{8} \right) \ln\frac{\Delta}{m_b}
    - \left( \frac14 \ln\frac{m_b}{\mu} - \frac{59}{96} \right) \beta_0
    + \left( 1 + \frac{\pi^2}{12} \right) C_F
    + \left( \frac16 - \frac{\pi^2}{24} \right) C_A \,, \nonumber\\
   k_3(\Delta)
   &=& \left( \frac{3\beta_0}{16} - \frac{C_F}{4} \right)
    \ln^2\frac{\Delta}{m_b}
    + \left[ \left( \frac14 \ln\frac{m_b}{\mu} - \frac{5}{96} \right) \beta_0 
    - \left( \frac12 + \frac{\pi^2}{12} \right) C_F 
    - \left( \frac16 - \frac{\pi^2}{24} \right) C_A \right]
    \ln\frac{\Delta}{m_b} \nonumber\\
   &&\mbox{}+ \left( \frac{5}{16} \ln\frac{m_b}{\mu} - \frac{5}{12}
    + \frac{\pi^2}{48} \right) \beta_0 
    - \left( \frac{11}{32} + \frac{\pi^2}{24} + \frac{\zeta_3}{4} \right) C_F
    + \left( \frac{13}{96} + \frac{\pi^2}{48}
    + \frac{\zeta_3}{8} \right) C_A \,, \nonumber\\
   k_4(\Delta)
   &=& \left( \frac{3C_F}{4} - \frac{3\beta_0}{16} \right)
    \ln\frac{\Delta}{m_b}
    - \left( \frac18 \ln\frac{m_b}{\mu} - \frac{41}{192} \right) \beta_0
    + \left( \frac78 + \frac{\pi^2}{24} \right) C_F
    + \left( \frac{1}{12} - \frac{\pi^2}{48} \right) C_A \,.
\end{eqnarray}}%
The terms of order $\alpha_s$ and $\beta_0\alpha_s^2$ in these expressions 
agree with those obtained in \cite{Ligeti:1999ea}. However, above we include
the complete set of two-loop corrections. Very recently, the dominant part of 
the $\bar B\to X_s\gamma$ photon spectrum has been calculated at 
$O(\alpha_s^2)$ \cite{Melnikov:2005bx}. Using the results of that paper to 
calculate the moments $M_n^{(0)}(\Delta)$, we find complete agreement with our 
expressions for the functions $k_1$ and $k_3$. We stress that both of these 
coefficients contain a contribution proportional to the combination 
$(2\gamma_1+\gamma_1^J)$ of the two-loop anomalous dimensions of the shape and 
jet functions. The agreement with \cite{Melnikov:2005bx} therefore serves as a 
test of the expressions for the anomalous dimensions collected in 
Appendix~A.2. 

Our expressions for the $\mu_\pi^2/\Delta^2$ power corrections to the moments 
are new, except for the tree-level contribution to $M_2^{(0)}(\Delta)$, which 
agrees with \cite{Kapustin:1995nr}. Prior to this work, power corrections 
proportional to the HQET parameter $\mu_\pi^2$ have never been computed 
beyond the tree approximation. Here, we have calculated the Wilson 
coefficients of these terms at two-loop order.

\subsection{Power corrections in \boldmath$1/m_b$\unboldmath}

The power-suppressed corrections to the moments $M_n(\Delta)$ can be separated 
into two classes: ``kinematic'' corrections of order $(\Delta/m_b)^k$, and 
``hadronic'' corrections involving non-pertur\-bative heavy-quark parameters. 
The kinematic corrections are known to $O(\alpha_s)$ in fixed-order 
perturbation theory, without scale separation and RG improvement. The Wilson 
coefficients of the operators contributing to the hadronic power corrections 
are known at tree level up to and including terms of order 
$\Lambda_{\rm QCD}^3$ in the heavy-quark expansion. The corresponding 
contributions to the moments are so small that radiative corrections to these
Wilson coefficients are unlikely to have any significant impact. The two types 
of corrections are linked to each other and should be combined consistently 
order by order in the $1/m_b$ expansion. When we will introduce heavy-quark 
parameters defined in physical renormalization schemes in 
Section~\ref{sec:RSchoice}, a reshuffling of terms between the kinematic and
hadronic corrections will take place.

We begin with a review of the kinematic power corrections. In fixed-order
perturbation theory, they are due to contributions from real-gluon emission 
graphs ($b\to s\gamma g$ at the parton level) that are phase-space suppressed 
in the endpoint region. The corresponding contributions to the partial decay 
rate can be included by adding the term
\begin{equation}\label{fijterms}
   \frac{C_F\alpha_s(\mu)}{4\pi} \sum_{i\le j} \mbox{Re}\,
   \frac{C_i^*(\mu_h)\,C_j(\mu_h)}{|C_7(\mu_h)|^2}\,3f_{ij}(\delta) 
\end{equation}
inside the curly brackets in the second line of (\ref{GOPE}). Here 
$\delta=\Delta/m_b$, and $i,j$ take the values $1,7,8$, corresponding to 
different operators in the effective weak Hamiltonian for 
$\bar B\to X_s\gamma$ decay. The coefficients 
$C_7\equiv C_{7\gamma}^{\rm eff}$ and $C_8\equiv C_{8g}^{\rm eff}$ are the 
``effective'' Wilson coefficients of the electro-magnetic and chromo-magnetic
dipole operators, while $C_1$ is the coefficient of the current--current 
operator $(\bar s c)_{V-A} (\bar c b)_{V-A}$. Operators other than these three 
can be safely neglected. The coefficients $C_i$ are evaluated at a hard scale
$\mu_h\sim m_b$. This will be implicitly understood in the equations below. On 
the other hand, given that the emitted gluon is part of the final-state 
hadronic jet $X_s$, an appropriate choice for the scale $\mu$ in the coupling 
constant in (\ref{fijterms}) is more likely to be one of the low scales 
$\mu_i$ or $\mu_0$. In our numerical analysis in Section~\ref{sec:numerics} we
set $\mu=\mu_i$ but vary both scales independently to be conservative.

In the Standard Model the Wilson coefficients are real, and we will thus drop 
the ``Re'' symbol  below. Explicit expressions for the kinematic functions 
$f_{ij}(\delta)$ can be found in \cite{Kagan:1998ym}. To first order in 
$\alpha_s$, the moments $M_n(\Delta)$ receive an additive contribution from 
the kinematic power corrections given by
\begin{equation}
   M_n(\Delta)\Big|_{\rm kin}
   = \frac{C_F\alpha_s(\mu)}{\pi} \sum_{i\le j} \frac{C_i C_j}{C_7^2}\,
   \frac34 \left[ f_{ij}(\delta) - n \int_0^1\!dy\,y^{n-1} f_{ij}(y\delta) 
   \right] ,
\end{equation}
where $C_i\equiv C_i(\mu_h)$. For the case $n=1$ the expressions in brackets 
were called $-d_{ij}(\delta)/\delta$ in \cite{Kagan:1998ym}, where explicit 
forms for these functions can be found. For our purposes it will be sufficient 
to expand the results in powers of $\delta$. The dominant contribution comes 
from the case where $i=j=7$, corresponding to weak decay mediated by the 
electro-magnetic dipole operator in the effective weak Hamiltonian. For this 
term, we also include the two-loop perturbative corrections, which can be 
extracted from the formulae given in \cite{Melnikov:2005bx}. Two-loop 
corrections involving other Wilson coefficients are presently unknown, but 
their effects on the moments are bound to be negligible.

Non-perturbative hadronic corrections of leading and subleading order were 
calculated in \cite{Kapustin:1995nr,Bauer:1997fe}, respectively. These authors 
find
\begin{eqnarray}
   M_1(\Delta)\Big|_{\rm hadr}
   &=& \frac{\lambda_1+3\lambda_2}{2m_b\Delta}
    + \frac{5\rho_1-21\rho_2}{6m_b^2\Delta}
    + \frac{{\cal T}_1+3{\cal T}_2+{\cal T}_3+3{\cal T}_4}{2m_b^2\Delta}
    + \dots \,, \nonumber\\
   M_2(\Delta)\Big|_{\rm hadr}
   &=& - \frac{\lambda_1}{3\Delta^2} - \frac{2\rho_1-3\rho_2}{3m_b\Delta^2}
    -  \frac{{\cal T}_1+3{\cal T}_2}{3m_b\Delta^2} + \dots \,,
\end{eqnarray}
where $\lambda_i$ \cite{Falk:1992wt}, $\rho_i$ \cite{Mannel:1994kv}, and 
${\cal T}_i$ \cite{Gremm:1996df} are hadronic matrix elements defined in HQET. 
Note that the leading term in the expression for the second moment is not 
power suppressed in $1/m_b$. It is already included in the leading-order
prediction for that moment in (\ref{M2lo}). In theoretical expressions for 
inclusive decay distributions, the parameters ${\cal T}_i$ always appear in 
the same combinations with $\lambda_1$ and $\lambda_2$, and it is thus 
convenient to absorb them into a redefinition of these parameters, such that
\begin{equation}
   \hat\lambda_1 = \lambda_1 + \frac{{\cal T}_1+3{\cal T}_2}{m_b} \,, \qquad
   \hat\lambda_2 = \lambda_2 + \frac{{\cal T}_3+3{\cal T}_4}{3m_b} \,.
\end{equation}
Then the only place where the ${\cal T}_i$ parameters appear is in 
spectroscopy. For instance, at tree level the spin splitting between the 
ground-state heavy mesons is given by \cite{Gremm:1996df}
\begin{equation}\label{BBstar}
   m_{B^*} - m_B = \frac{2\hat\lambda_2}{m_b}
   - \frac{\rho_2-{\cal T}_2+\frac23\,{\cal T}_3+{\cal T}_4}{m_b^2} + \dots \,,
\end{equation}
which in essence means that there is an uncertainty of relative order 
$\Lambda_{\rm QCD}/m_b$ in the determination of the parameter $\hat\lambda_2$.
For the purposes of the present work we adopt the conventions introduced in  
\cite{Bigi:1994ga}, which are such that
\begin{equation}
   \mu_\pi^2 = - \hat\lambda_1 \,, \qquad
   \mu_G^2 = 3\hat\lambda_2 \,, \qquad
   \rho_D^3 = \rho_1 \,, \qquad
   \rho_{LS}^3 = 3\rho_2 \,.
\end{equation}
For the time being all definitions still refer to the pole scheme. 

Combining the contributions from kinematic and hadronic power corrections, we
find that the first-order corrections to the moments are
\begin{eqnarray}\label{M11}
   M_1^{(1)}(\Delta)
   &=& \frac{\Delta}{m_b}\,\Bigg\{
    \frac{\mu_G^2-\mu_\pi^2}{2\Delta^2}
    + \frac{C_F\alpha_s(\mu)}{\pi}\,\Bigg[
    1 - \frac12 \ln\frac{\Delta}{m_b}
    + \frac{\alpha_s(\mu)}{\pi}\,l_1(\Delta)
    + \frac29\,\frac{C_1^2}{C_7^2}\,g_2(z)
    \nonumber\\
   &&\mbox{}+ \frac{5}{12}\,\frac{C_8}{C_7}
    - \frac13 \left( \frac{C_1}{C_7} - \frac{C_1 C_8}{3C_7^2} \right) g_1(z)
    + \frac{1}{18}\,\frac{C_8^2}{C_7^2} \left( \ln\frac{m_b}{m_s} - 1
    + \frac12\ln\frac{\Delta}{m_b} \right) \Bigg] \Bigg\} \,, \nonumber\\
   M_2^{(1)}(\Delta)
   &=& \frac{\Delta}{m_b}\,\Bigg\{
    - \frac{2\rho_D^3-\rho_{LS}^3}{3\Delta^3}
    + \frac{C_F\alpha_s(\mu)}{\pi}\,\Bigg[
    \frac{11}{18} - \frac13 \ln\frac{\Delta}{m_b}
    + \frac{\alpha_s(\mu)}{\pi}\,l_2(\Delta)
    + \frac{4}{27}\,\frac{C_1^2}{C_7^2}\,g_2(z) \nonumber\\
   &&\mbox{}+ \frac{5}{18}\,\frac{C_8}{C_7} 
    - \frac29 \left( \frac{C_1}{C_7} - \frac{C_1 C_8}{3C_7^2} \right) g_1(z)
    + \frac{1}{27}\,\frac{C_8^2}{C_7^2} \left( \ln\frac{m_b}{m_s}
    - \frac{11}{12} + \frac12\ln\frac{\Delta}{m_b} \right) \Bigg] \Bigg\} \,,
\end{eqnarray}
where we have ordered the various contributions in magnitude. The combined 
result for the second-order power correction to the first moment reads
\begin{eqnarray}\label{M12}
   M_1^{(2)}(\Delta)
   &=& \left( \frac{\Delta}{m_b} \right)^2 \Bigg\{
    \frac{5\rho_D^3-7\rho_{LS}^3}{6\Delta^3}
    + \frac{C_F\alpha_s(\mu)}{\pi}\,\Bigg[
    \frac{7}{36} + \frac16 \ln\frac{\Delta}{m_b}
    + \frac{\alpha_s(\mu)}{\pi}\,l_3(\Delta) \nonumber\\
   &&\mbox{}+ \frac{C_8}{C_7} \left( - \frac{5}{27}
    + \frac29 \ln\frac{\Delta}{m_b} \right)
    + \frac29 \left( \frac{C_1}{C_7} - \frac{C_1 C_8}{3C_7^2} \right) g_3(z)
    \nonumber\\
   &&\mbox{}+ \frac{1}{27}\,\frac{C_8^2}{C_7^2} \left( \ln\frac{m_b}{m_s}
    - \frac{11}{12} + \frac12\ln\frac{\Delta}{m_b} \right) \Bigg] \Bigg\} \,.
\end{eqnarray}
In these expressions,
\begin{eqnarray}
   g_1(z) &=& \int_0^1\!dx\,x\,\mbox{Re} \left[\,
    \frac{z}{x}\,G\!\left(\frac{x}{z}\right) + \frac12 \,\right] , \nonumber\\
   g_2(z) &=& \int_0^1\!dx\,(1-x) \left|\,\frac{z}{x}\,
    G\!\left(\frac{x}{z}\right) + \frac12\,\right|^2 , \nonumber\\
   g_3(z) &=& \mbox{Re} \left[ z\,G\!\left(\frac{1}{z}\right) + \frac12
    \,\right] ,
\end{eqnarray}
with
\begin{equation}
   G(t) = \left\{ \begin{array}{ll}
    -2\arctan^2\!\sqrt{t/(4-t)} & ;~ t<4 \,, \\[0.1cm]
    2 \left( \ln\!\Big[(\sqrt{t}+\sqrt{t-4})/2\Big]
    - \displaystyle\frac{i\pi}{2} \right)^2 & ;~ t\ge 4 \,,
   \end{array} \right.
\end{equation}
are functions of the mass ratio $z=(m_c/m_b)^2$, which arise from charm-penguin
loop diagrams. The logarithms of the quark-mass ratio $m_b/m_s$ arise due to a
collinear singularity in the process $b\to sg$ mediated by the operator 
$Q_{8g}$, followed by photon emission off the strange quark. This contribution 
is so small that a more careful treatment of these logarithms is not 
necessary. The $O(\alpha_s^2)$ corrections to the electro-magnetic dipole 
contributions in (\ref{M11}) and (\ref{M12}) are encoded in the coefficients
{\small
\begin{eqnarray}
   l_1(\Delta)
   &=& \left( \frac{3\beta_0}{16} - C_F \right) \ln^2\frac{\Delta}{m_b}
    + \left[
    \left( \frac14 \ln\frac{m_b}{\mu} - \frac{11}{24} \right) \beta_0 
    + \left( \frac52 - \frac{\pi^2}{8} \right) C_F 
    - \left( \frac{11}{48} - \frac{\pi^2}{16} \right) C_A \right]
    \ln\frac{\Delta}{m_b} \nonumber\\
   &&\mbox{}- \left( \frac12 \ln\frac{m_b}{\mu} - \frac{43}{96}
    - \frac{\pi^2}{48} \right) \beta_0 
    + \left( \frac{149}{48} + \frac{5\pi^2}{36} - \frac98\,\zeta_3 \right) C_F
    - \left( \frac{5}{16} + \frac{35\pi^2}{288} - \frac{9}{16}\,\zeta_3
    \right) C_A \,, \nonumber\\
   l_2(\Delta)
   &=& \left( \frac{\beta_0}{8} - \frac{4C_F}{9} \right)
    \ln^2\frac{\Delta}{m_b}
    + \left[
    \left( \frac16 \ln\frac{m_b}{\mu} - \frac{19}{72}  \right) \beta_0 
    + \left( \frac{19}{18} - \frac{\pi^2}{12} \right) C_F 
    - \left( \frac{11}{72} - \frac{\pi^2}{24} \right) C_A \right]
    \ln\frac{\Delta}{m_b} \nonumber\\
   &&\mbox{}- \left(  \frac{11}{36} \ln\frac{m_b}{\mu} - \frac{101}{432}
    - \frac{\pi^2}{72} \right) \beta_0 
    + \left( \frac{977}{432} + \frac{17\pi^2}{216} - \frac34\,\zeta_3 \right)
    C_F
    - \left( \frac{101}{432} + \frac{2\pi^2}{27} - \frac38\,\zeta_3 \right)
    C_A \,, \nonumber\\
   l_3(\Delta)
   &=& \frac{C_F}{6} \ln^3\frac{\Delta}{m_b}
    + \left( \frac{2C_F}{9} + \frac{C_A}{16} - \frac{\beta_0}{16} \right)
    \ln^2\frac{\Delta}{m_b} \nonumber\\
   &&\mbox{}+ \left[ - \left( \frac{1}{12} \ln\frac{m_b}{\mu}
    - \frac{19}{144} \right) \beta_0 
    + \left( \frac{293}{144} + \frac{\pi^2}{36} \right) C_F 
    + \left( \frac{11}{36} - \frac{\pi^2}{72} \right) C_A \right]
    \ln\frac{\Delta}{m_b} \\
   &&\mbox{}- \!\left( \frac{7}{72} \ln\frac{m_b}{\mu} + \frac{65}{864}
    + \frac{\pi^2}{144} \right) \beta_0 
    - \!\left( \frac{1451}{864} + \frac{11\pi^2}{216} - \frac{5}{12}\,\zeta_3
    \right) C_F
    + \!\left( \frac{149}{216} - \frac{13\pi^2}{432} - \frac{5}{24}\,\zeta_3
    \right) C_A \,. \nonumber
\end{eqnarray}}%

This completes our compilation of theoretical formulae for the moments in the
pole scheme. The first moment, $M_1(\Delta)$, and with it the average photon
energy $\langle E_\gamma\rangle$, can be calculated including first- and
second-order power corrections in the $1/m_b$ expansion. For the second 
moment, $M_2(\Delta)$, and hence for the variance $\sigma_E^2$, only 
first-order power corrections are available at present.

\subsection{Elimination of pole-scheme parameters}
\label{sec:RSchoice}

It is well known that heavy-quark parameters defined in the pole scheme suffer 
from infra-red renormalon ambiguities 
\cite{Bigi:1994em,Beneke:1994sw,Martinelli:1995zw,Neubert:1996zy}. As a 
result, the perturbative expansion for the moments presented in the previous 
section would not be well behaved. It is necessary to replace the pole mass 
$m_b$ and other HQET parameters such as $\mu_\pi^2$ in favor of some physical, 
short-distance quantities. For our purposes, the ``shape-function scheme'' 
proposed in \cite{Bosch:2004th} provides for a particularly suitable 
definition of the heavy-quark mass and kinetic energy. In this scheme, 
low-scale subtracted HQET parameters are defined via the moments of the 
renormalized shape function, regularized with a hard Wilsonian cutoff 
$\mu_f\gg\Lambda_{\rm QCD}$. In addition to their dependence on the cutoff, 
the HQET parameters depend on the scale $\mu$ at which the shape function is 
renormalized. For simplicity, we adopt the ``diagonal'' scale choice 
$\mu=\mu_f$. The conventional choice for the subtraction scale is 
$\mu_f=1.5$\,GeV.

At two-loop order, the heavy-quark parameters $m_b(\mu_f)$ and 
$\mu_\pi^2(\mu_f)$ defined in the shape-function scheme are related to the 
pole-scheme parameters by \cite{Neubert:2004sp,Lange:2005yw}
\begin{eqnarray}\label{SFscheme}
   m_b \Big|_{\rm pole}
   &=& m_b(\mu_f) + \mu_f\,\frac{C_F\alpha_s(\mu)}{\pi}\,
    \Bigg\{ 1 + \frac{\alpha_s(\mu)}{\pi} \Bigg[
    \left( \frac12 \ln\frac{\mu}{\mu_f} + \frac{47}{36} \right) \beta_0
    + \left( \frac{23}{18} - \frac{\pi^2}{12} - \frac94\,\zeta_3
    \right) C_A \nonumber\\
   &&\qquad\qquad\mbox{}- 
    \left( 8 - \frac{\pi^2}{3} - 4\zeta_3 \right) C_F \Bigg]
    \Bigg\} \nonumber\\
   &&\mbox{}+ \frac{\mu_\pi^2(\mu_f)}{\mu_f}\,C_F
    \left( \frac{\alpha_s(\mu)}{\pi} \right)^2
    \left[ - \frac{5\beta_0}{108}
    - \left( \frac{17}{54} - \frac34\,\zeta_3 \right) C_A
    + \left( 1 - \frac43\,\zeta_3 \right) C_F \right] \nonumber\\ 
   &&\mbox{}- \frac{\mu_f^2}{2m_b(\mu_f)}\,
    C_F \left( \frac{\alpha_s(\mu)}{\pi} \right)^2
    \left[ - \frac{7\beta_0}{12}
    - \left( \frac{17}{12} - \frac{27}{8}\,\zeta_3 \right) C_A
    + \left( \frac{27}{4} - 6\zeta_3 \right) C_F \right] \nonumber\\
   &&\mbox{}+ \frac{\mu_G^2-\mu_\pi^2(\mu_f)}{2m_b(\mu_f)} + \dots \,, 
    \nonumber\\
   \mu_\pi^2 \Big|_{\rm pole}
   &=& \mu_\pi^2(\mu_f)\,\Bigg\{ 1 - \frac{C_F\alpha_s(\mu)}{2\pi}
    + C_F \left( \frac{\alpha_s(\mu)}{\pi} \right)^2
    \Bigg[ - \left( \frac14 \ln\frac{\mu}{\mu_f} - \frac18 \right) \beta_0
     \nonumber\\
   &&\qquad\qquad\mbox{}+
    \left( \frac54 + \frac{\pi^2}{24} - \frac{27}{8}\,\zeta_3
    \right) C_A
    - \left( \frac{13}{4} + \frac{\pi^2}{6} - 6\zeta_3 \right) C_F \Bigg]
    \Bigg\} \nonumber\\
   &&\mbox{}+ \mu_f^2\,C_F \left( \frac{\alpha_s(\mu)}{\pi} \right)^2
    \left[ - \frac{7\beta_0}{12}
    - \left( \frac{17}{12} - \frac{27}{8}\,\zeta_3 \right) C_A
    + \left( \frac{27}{4} - 6\zeta_3 \right) C_F \right] + \dots \,.
\end{eqnarray}
The analogous replacement rule for the parameter $\Delta=m_b-2E_0$ follows 
from its definition. When these expressions are substituted for the 
pole-scheme parameters in the formulae for the moments, the results must be 
reexpanded consistently to the desired order in $\alpha_s$ and $1/m_b$. In the 
process, one finds that for the spectral moments in (\ref{physmoments}), the 
$1/m_b$-suppressed term in the replacement rule for the pole mass cancels 
against the contribution proportional to $(\mu_G^2-\mu_\pi^2)$ from 
$M_1^{(1)}$ in (\ref{M11}). In the shape-function scheme, the predictions for 
$\langle E_\gamma\rangle$ and $\sigma_E^2$ are therefore independent of the 
parameter $\mu_G^2$.

The arbitrary renormalization point $\mu$ of the running coupling 
$\alpha_s(\mu)$ in the relations (\ref{SFscheme}) will be set equal to the 
matching scale $\mu_0$ in all factorized, leading-power contributions. For the 
power corrections, where no scale separation is available, $\mu$ will be 
identified with the common renormalization scale in (\ref{M11}) and 
(\ref{M12}). The subtraction scale $\mu_f$ of the shape-function scheme is set 
by the upper limit on $\omega$ in shape-function integrals of the type 
(\ref{contour}). For the case of the $\bar B\to X_s\gamma$ moments this yields 
$\mu_f\sim\Delta\sim\mu_0$. Note that this implies a reshuffling between 
perturbative and non-perturbative terms at any given order in the $1/m_b$ 
expansion. For instance, eliminating the pole-scheme kinetic-energy parameter 
from the $\mu_\pi^2/\Delta^2$ power corrections in (\ref{Mnmaster}) adds terms 
of order $[\alpha_s(\mu_0)]^2$ to the leading-power contributions. As 
mentioned above, we will adopt the conventional choice $\mu_f=1.5$\,GeV in our 
numerical analysis, which is slightly larger than (but of the same magnitude 
as) the actual value $\Delta\approx 1$\,GeV.

While it is natural to use the shape-function scheme for analyses of inclusive 
$B$ decays, this is by no means mandatory. In Appendix~A.3, we present the 
corresponding replacement rules for the kinetic scheme introduced in 
\cite{Benson:2003kp}. We do not explore alternative short-distance definitions 
of the $b$-quark mass, such as the ``$1S$ mass'' \cite{Hoang:1998hm} or the 
``potential-subtracted mass'' \cite{Beneke:1998rk}. The reason is that no 
physical definition of the kinetic-energy parameter $\mu_\pi^2$ has been 
provided in these schemes, so they remove the renormalon problem only 
partially.

\section{Boost to the \boldmath$\Upsilon(4S)$ rest frame\unboldmath}

Theoretical calculations of $B$-meson decay distributions are easiest to 
perform in the rest frame of the heavy meson. In all our results so far, 
$E_\gamma$ denotes the photon energy measured in that frame. Existing 
experimental studies of the decay $\bar B\to X_s\gamma$ 
\cite{Chen:2001fj,Koppenburg:2004fz,Aubert:2005cu}, however, measure the 
photon energy in a different frame. At an $e^+ e^-$ $B$-factory, pairs of 
$B\bar B$ mesons are produced on the $\Upsilon(4S)$ resonance peak. For CLEO, 
the $\Upsilon(4S)$ rest frame coincides with the laboratory system, whereas 
for BaBar and Belle the $\Upsilon(4S)$ rest frame can be constructed knowing 
the (asymmetric) energies of the electron and positron beams. In either case, 
the photon energy is measured in the $\Upsilon(4S)$ rest frame, in which the 
$B$ mesons have a small velocity 
\begin{equation}
   \beta = \sqrt{1 - \frac{4m_B^2}{m_{\Upsilon(4S)}^2}} \approx 0.064 \,.
\end{equation}
Below, we work out in detail how the properties of the photon spectrum and its 
moments are affected by boosting from the $B$-meson rest frame to the rest 
frame of the $\Upsilon(4S)$ resonance, following \cite{Kagan:1998ym}. While 
this would be straightforward for moments of the entire spectrum, the presence 
of the cut leads to non-trivial complications. Note that future measurements 
of $\bar B\to X_s\gamma$ decay spectra relying on the full-reconstruction 
technique could reconstruct the $B$-meson rest frame and thus directly measure 
the spectrum in that reference system.

We denote quantities in the $\Upsilon(4S)$ rest frame with a prime. Let
$k^\mu=E_\gamma(1,\bm{n})$ be the 4-vector of the photon in the $B$-meson rest
frame. In the $\Upsilon(4S)$ system the $B$ meson moves with velocity 
$\bm{\beta}$, and the photon energy is Doppler-shifted by an amount
\begin{equation}
   E_\gamma' = E_\gamma\,\frac{1 - \bm{\beta}\cdot\bm{n}}{\sqrt{1-\beta^2}}
   = E_\gamma\,\frac{1 - \beta\cos\theta}{\sqrt{1-\beta^2}} \,,
\end{equation}
where we have assumed without loss of generality that $\bm{\beta}$ points in 
the $z$-direction. The photon spectrum $dN/dE_\gamma'$ in the $\Upsilon(4S)$ 
rest frame can be obtained in terms of the spectrum $dN/dE_\gamma$ in the 
$B$-meson rest frame by evaluating
\begin{eqnarray}
   \frac{dN}{dE_\gamma'}
   &=& \int \frac{d\phi\,d\cos\theta}{4\pi}
    \int\!dE_\gamma\,\frac{dN}{dE_\gamma}\,
    \delta\bigg( E_\gamma' - E_\gamma\,
    \frac{1 - \beta\cos\theta}{\sqrt{1-\beta^2}} \bigg) \nonumber\\
   &=& \frac{1}{\beta_+ - \beta_-}\!\!
    \int\limits_{\beta_- E_\gamma'}^{{\rm min}
    (\beta_+ E_\gamma',E_\gamma^{\rm max})}\!\!\!dE_\gamma\,
    \frac{1}{E_\gamma}\,\frac{dN}{dE_\gamma} \,,
\end{eqnarray}
where $\beta_\pm=\sqrt{(1\pm\beta)/(1\mp\beta)}$. This reproduces eq.~(B.1) in 
\cite{Kagan:1998ym}. 

It is now straightforward to compute the effect of the boost on moments of the 
photon spectrum. We obtain
\begin{eqnarray}\label{momentsboost}
   \int\limits_{E_0}^{E_\gamma^{\prime\,{\rm max}}}\!\!\!dE_\gamma'\,
   (E_\gamma')^n\,\frac{dN}{dE_\gamma'}
   &=& \frac{\beta_+^{n+1} - \beta_-^{n+1}}{(n+1)(\beta_+ - \beta_-)}
    \int\limits_{E_0}^{E_\gamma^{\rm max}}\!\!\!dE_\gamma\,
    (E_\gamma)^n\,\frac{dN}{dE_\gamma} \\
   &&\hspace{-3.5cm}\mbox{}+ \frac{1}{(n+1)(\beta_+ - \beta_-)} \left[ ~
    \int\limits_{\beta_- E_0}^{E_0}\!\!\!dE_\gamma\,\frac{1}{E_\gamma}\,
    \frac{dN}{dE_\gamma} \left[ (\beta_+ E_\gamma)^{n+1} - (E_0)^{n+1} \right]
    - (\beta_+\leftrightarrow\beta_-) ~ \right] , \nonumber
\end{eqnarray}
where $E_\gamma^{\prime\,{\rm max}}=\beta_+ E_\gamma^{\rm max}$, and the same 
value of the cutoff $E_0$ must be used on both sides of the equation. In order 
to illustrate the effect of the boost on the moments, we use the theoretical 
description of the $\bar B\to X_s\gamma$ photon spectrum at NLO presented in 
\cite{Lange:2005yw} to generate the distribution $dN/dE_\gamma$ in the 
$B$-meson rest frame. We adopt the default choices for all parameters and use 
two models of the shape function, corresponding to heavy-quark parameters 
$m_b(\mu_f)=4.61$\,GeV, $\mu_\pi^2(\mu_f)=0.2$\,GeV$^2$ (model~1) and 
$m_b(\mu_f)=4.55$\,GeV, $\mu_\pi^2(\mu_f)=0.3$\,GeV$^2$ (model~2) at 
$\mu_f=1.5$\,GeV. In Table~\ref{tab:boost}, we compare the results for the 
average photon energy and the variance of the spectrum in the $B$-meson and 
$\Upsilon(4S)$ rest frames. Numerically, it turns out that the terms 
containing the cutoff $E_0$ in (\ref{momentsboost}) have a small effect. 
Keeping only the contribution in the first line of that equation 
(corresponding to the limit where $E_0=0$) yields the simple relations
\begin{eqnarray}
   \langle E_\gamma'\rangle - \langle E_\gamma\rangle
   &=& \left( \frac{1}{\sqrt{1-\beta^2}} - 1 \right) \langle E_\gamma\rangle
    \approx 0.005\,\mbox{GeV} \,, \nonumber\\
   \sigma_{E'}^2 - \sigma_E^2
   &=& \frac{\beta^2}{3(1-\beta^2)} \left[ \langle E_\gamma\rangle^2
    + 4\sigma_E^2 \right]
    \approx 0.007\,\mbox{GeV}^2 \,.
\end{eqnarray}
This explains to a large extent the shifts seen in the table.

\begin{table}
\centerline{\parbox{14cm}{\caption{\label{tab:boost}
Predictions for the average photon energy and the variance of the photon 
spectrum in the $B$-meson and $\Upsilon(4S)$ rest frames, for different values 
of the cutoff $E_0$. In each case, the first line refers to shape-function 
model~1, the second line to model~2. See text for explanation.}}}
\vspace{0.1cm}
\begin{center}
\begin{tabular}{|c|cc|cc|}
\hline\hline
 & \multicolumn{2}{c|}{$\langle E_\gamma\rangle$~[GeV]}
 & \multicolumn{2}{c|}{$\sigma_E^2$~[$10^{-2}\,\mbox{GeV}^2$]} \\
$E_0$~[GeV] & $B$ frame & $\Upsilon(4S)$ frame & $B$ frame
 & $\Upsilon(4S)$ frame \\
\hline
1.8 & 2.311 & 2.317 & 3.10 & 3.76 \\
    & 2.286 & 2.292 & 3.73 & 4.35 \\
\hline
1.9 & 2.325 & 2.331 & 2.54 & 3.17 \\
    & 2.305 & 2.313 & 3.01 & 3.59 \\
\hline
2.0 & 2.341 & 2.350 & 1.99 & 2.59 \\
    & 2.329 & 2.338 & 2.31 & 2.87 \\
\hline\hline
\end{tabular}
\end{center}
\end{table}

\section{Numerical analysis}
\label{sec:numerics}

The theoretical expressions for the spectral moments depend on several input 
parameters, whose values are summarized in Table~\ref{tab:inputs}. The 
relevant hadronic parameters are $\mu_\pi^2$, $\rho_D^3$, $\rho_{LS}^3$, and 
$\mu_G^2$ if the kinetic scheme is used instead of the shape-function scheme. 
As mentioned earlier, at tree level the results for the spectral moments are 
$\langle E_\gamma\rangle=m_b/2+\dots$ and $\sigma_E^2=\mu_\pi^2/12+\dots$, and 
our goal will be to determine the parameters $(m_b,\mu_\pi^2)$ from a fit to 
experimental data. The sensitivity of the moments to other hadronic parameters 
is very weak (see below), so it is safe to use them as fixed inputs in the 
fit. Following \cite{Gambino:2004qm}, we define the quantities $\rho_D^3$ and 
$\rho_{LS}^3$ in the pole scheme, which is justified given the smallness of 
their contributions to the moments. We use as inputs the values 
$\rho_D^3=(0.195\pm 0.029)$\,GeV$^3$ and 
$\rho_{LS}^3=(-0.085\pm 0.082)$\,GeV$^3$ extracted from a global fit to 
$\bar B\to X_c\,l^-\bar\nu$ moments in the kinetic scheme \cite{Aubert:2004aw}
and convert them to the pole scheme by subtracting 
$[2C_F\alpha_s(2\mu_f)/3\pi]\,\mu_f^3\approx 0.09$\,GeV$^3$ (at 
$\mu_f=1$\,GeV) from $\rho_D^3$ \cite{Benson:2003kp}. We inflate the error on 
$\rho_D^3$ from 0.03 to 0.05\,GeV$^3$ to be conservative. The resulting values 
are consistent with, but more accurate than, the theoretical estimates 
$\rho_D^3=(0.1\pm 0.1)$\,GeV$^3$ and $\rho_{LS}^3=(-0.15\pm 0.10)$\,GeV$^3$ 
given in \cite{Benson:2003kp,Bigi:1997fj}. The value for $\rho_D^3$ is also in 
agreement with early estimates using the vacuum-insertion approximation, which 
gave $\rho_D^3\approx(2\pi\alpha_s/9)\,f_B^2 m_B\approx 0.1$\,GeV$^3$
\cite{Bigi:1993ex,Mannel:1994pm}. The value of $\mu_G^2$ quoted in the table 
is derived using (\ref{BBstar}) and assigning an error for possible $1/m_b$ 
corrections \cite{Bigi:1997fj}. The formulae for the power corrections to the 
moments involve numerically small terms depending on the quark-mass ratios 
$z=(m_c/m_b)^2$ and $m_s/m_b$, for which we adopt values consistent with 
\cite{Neubert:2004dd}. Throughout, we use the three-loop running coupling 
constant, matched to a four-flavor theory at 
$\mu=\overline{m}_b(\overline{m}_b)=4.25$\,GeV. 

\begin{table}
\centerline{\parbox{14cm}{\caption{\label{tab:inputs}
Compilation of input parameters.}}}
\vspace{0.1cm}
\begin{center}
\begin{tabular}{|c|c|}
\hline\hline
Parameter & Value \\
\hline
$\rho_D^3$ & $(0.11\pm 0.05)$\,GeV$^3$ \\
$\rho_{LS}^3$ & $(-0.09\pm 0.08)$\,GeV$^3$ \\
$\mu_G^2$ & $(0.35\pm 0.07)$\,GeV$^2$ \\
$m_c/m_b$ & $0.222\pm 0.030$ \\
$m_s/m_b$ & 0.02 \\
$\alpha_s(m_Z)$ & 0.1187 \\
\hline\hline
\end{tabular}
\end{center}
\end{table}

Because of the truncation of perturbation theory, our results are sensitive to 
the choice of the various factorization scales. This sensitivity can be taken 
as an estimator of the residual perturbative uncertainty. In the resummed 
expressions obtained in RG-improved perturbation theory, we vary the three 
matching scales by a factor between $1/\sqrt2$ and $\sqrt2$ about their 
default values $\mu_h^{\rm def}=m_b$, $\mu_i^{\rm def}=\sqrt{m_b\Delta}$, and 
$\mu_0^{\rm def}=\Delta$, using $m_b=4.65$\,GeV as a reference value. The 
scale $\mu$ in the expressions for the power corrections is set equal to 
$\mu_i$ but varied independently. Thus, for $E_0=1.8$\,GeV we vary 
$\mu_h\in[3.29,6.58]$\,GeV, $\mu_i,\mu\in[1.56,3.12]$\,GeV, and 
$\mu_0\in[0.74,1.48]$\,GeV. Together this covers a conservative range of 
scales. When quoting results in fixed-order perturbation theory (in which all 
scales are set equal to $\mu$), we vary $\mu$ between 1 and 5\,GeV.

\subsection{Predictions for the moments of the photon spectrum}

We begin by presenting predictions for the average photon energy and the 
variance of the photon spectrum, including a detailed account of theoretical 
uncertainties. We define the heavy-quark parameters $m_b$ and $\mu_\pi^2$ in 
the shape-function scheme at a subtraction point $\mu_f=1.5$\,GeV. For 
reference, we recall that the values for these parameters extracted from a 
global fit to $\bar B\to X_c\,l^-\bar\nu$ moments are 
$m_b=(4.61\pm 0.08)$\,GeV and $\mu_\pi^2=(0.15\pm 0.07)$\,GeV$^2$ 
\cite{Neubert:2004sp}, where we account for the small $1/m_b$ corrections in 
the relation for the pole mass in (\ref{SFscheme}), which were not included in 
that paper. For the purpose of illustration, we use the central values of 
these parameters for the following discussion.

\begin{table}
\centerline{\parbox{14cm}{\caption{\label{tab:benchmark}
Predictions and error analysis for the first two moments of the photon 
spectrum, defined with a cutoff $E_0=1.8$\,GeV. The parameters $m_b=4.61$\,GeV 
and $\mu_\pi^2=0.15$\,GeV$^2$ are kept fixed. In each column, the upper 
(lower) error indicates the variation obtained by increasing (decreasing) a 
given input parameter.}}}
\vspace{0.1cm}
\begin{center}
\begin{tabular}{|c|c|cccc|ccc|}
\hline\hline
Moment & Value & $\mu_0$ & $\mu_i$ & $\mu_h$ & $\mu$ & $m_c/m_b$ & $\rho_D^3$
 & $\rho_{LS}^3$ \\
\hline
$\langle E_\gamma\rangle$~[$10^{-3}$\,GeV]
 & 2287 & ${}_{\,+22}^{\,-\phantom{1}1}$ & ${}_{\,+\phantom{1}7}^{\,-11}$
 & $\mp 1$ & ${}_{\,-7}^{\,+5}$ & ${}_{\,-1}^{\,+2}$ & $\mp 1$ & $\pm 2$ \\
\hline
$\sigma_E^2$~[$10^{-4}$\,GeV$^2$]
 & 334 & ${}_{\,-84}^{\,+15}$ & ${}_{\,+47}^{\,-26}$ & $\pm 2$
 & ${}_{\,+12}^{\,-11}$ & ${}_{\,+3}^{\,-4}$ & $\mp 18$ & $\pm 14$ \\
\hline\hline
\end{tabular}
\end{center}
\end{table}

The predictions for the average photon energy and variance are obtained using 
the relations in (\ref{physmoments}). In calculating 
$\langle E_\gamma\rangle$, we include both first- and second-order power 
corrections to the moment $M_1$, as given in Section~\ref{sec:moments}. When 
calculating the variance, we use the second relation in (\ref{physmoments}) 
and compute, for consistency, all quantities ($M_1$, $M_2$, and 
$\langle E_\gamma\rangle$) including first-order power corrections. 
Table~\ref{tab:benchmark} shows our results for the case $E_0=1.8$\,GeV, 
corresponding to the lowest value of the cutoff that has so far been achieved 
experimentally \cite{Koppenburg:2004fz}. As expected, the perturbative 
uncertainties are larger the smaller the relevant matching scales are, but 
they remain under good control even for the lowest scale $\mu_0$. The 
sensitivity to other input parameters, and in particular to the hadronic 
quantities $\rho_D^3$ and $\rho_{LS}^3$, is very small. The convergence of the 
heavy-quark expansion is good for both moments. The first-order power 
correction to the average photon energy lowers the value of 
$\langle E_\gamma\rangle$ by about 54\,MeV, corresponding to an 11\% reduction 
of the difference $(\langle E_\gamma\rangle-E_0)$, which is the relevant 
quantity to compare with. The impact of the second-order power correction is 
negligible ($+2$\,MeV, corresponding to a 0.4\% increase). The first-order 
power correction to the second moment is larger and constitutes about 27\% of 
the total value. The main effect is due to the first-order correction to 
$\langle E_\gamma\rangle$, which enters via the second term in the relation 
for the variance in (\ref{physmoments}).

\begin{figure}[t]
\begin{center}
\epsfig{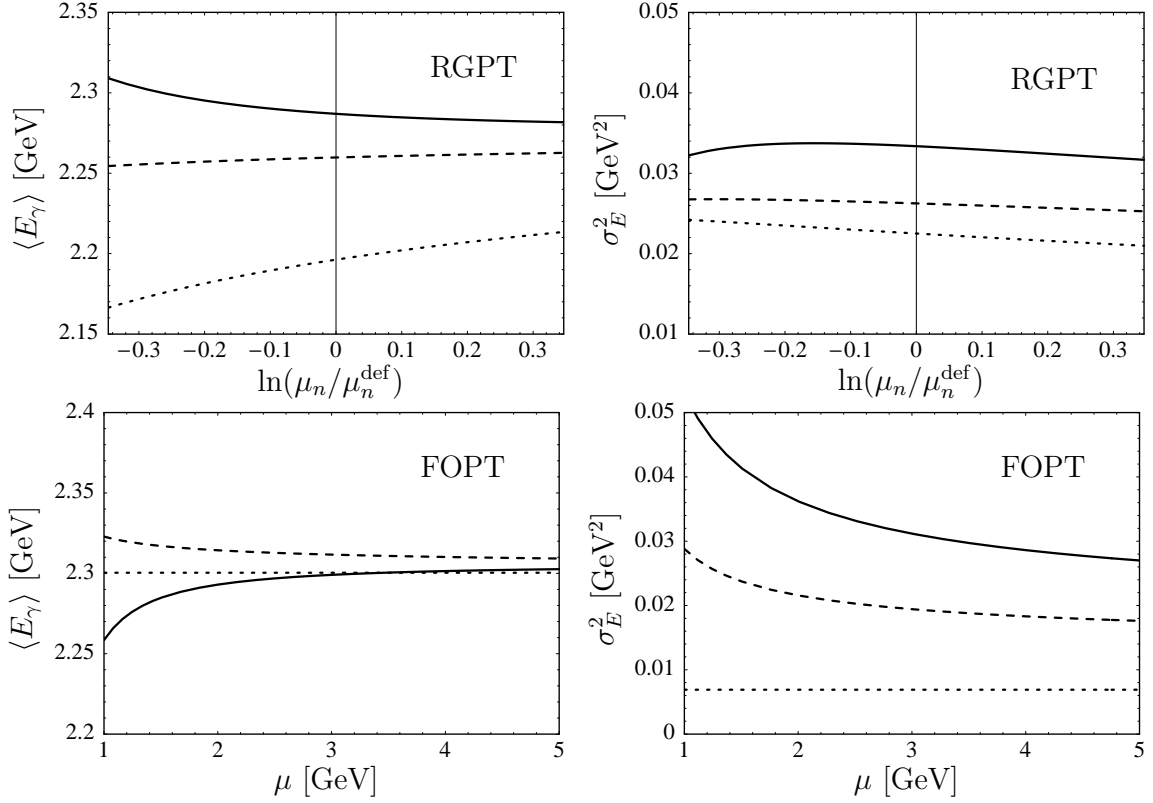}
\end{center}
\vspace{-0.2cm}
\centerline{\parbox{15cm}{\caption{\label{fig:mudep}
Scale dependence of the predictions for the first two moments of the photon 
spectrum, defined with a cut at $E_0=1.8$\,GeV in the $B$-meson rest frame, in 
RG-improved perturbation theory (RGPT, top) and fixed-order perturbation 
theory (FOPT, bottom). Solid, dashed, and dotted lines refer to the NNLO, NLO, 
and LO approximations. All parameters are set to their default values.}}}
\end{figure}

Next, we study the behavior of the perturbative expansions of the moments and 
explore to what extent it is improved by using scale separation and RG 
improvement, which is one of the main new features of our approach. 
Figure~\ref{fig:mudep} shows our predictions for the average photon energy and 
variance, using both RG-improved and fixed-order perturbation theory. In each 
plot the solid, dashed, and dotted curves correspond to the NNLO 
($\sim\!\!\alpha_s^2$), NLO ($\sim\!\!\alpha_s$), and LO 
($\sim\!\!\alpha_s^0$) approximations, respectively. In the factorized 
expressions, the four relevant scales ($\mu_n=\mu_h,\mu_i,\mu_0,\mu$) are 
varied simultaneously (and in a correlated way) about their default values 
($\mu_n^{\rm def}$). We observe an excellent stability of the RG-improved 
results under scale variation, both at NLO and NNLO. In the case of 
fixed-order perturbation theory, on the other hand, the results obtained at
NNLO are less stable than those obtained at NLO (the tree-level LO results are 
trivially scale independent in fixed-order calculations). While the absolute 
variations are still modest for the average photon energy, they are quite 
large for the case of the variance. We conclude that a proper scale separation
is important for obtaining reliable perturbative predictions for the moments.

\begin{figure}[t]
\begin{center}
\epsfig{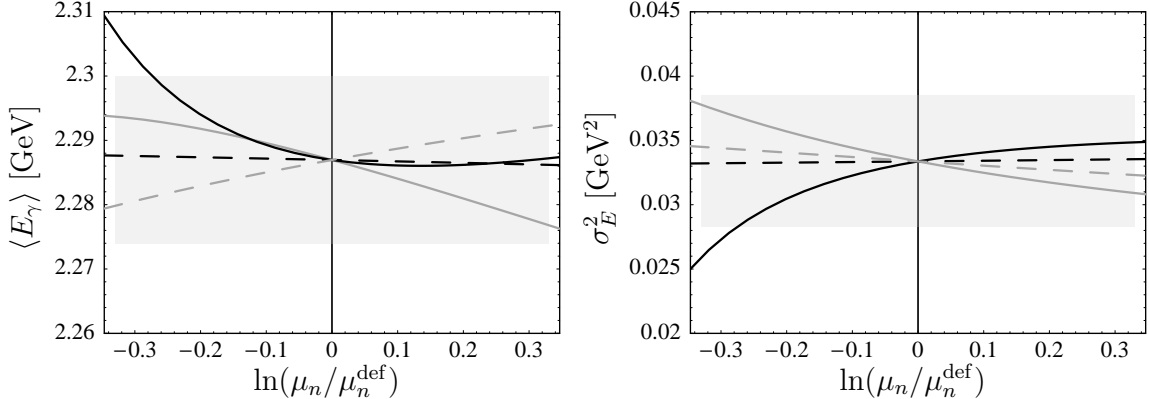}
\end{center}
\vspace{-0.2cm}
\centerline{\parbox{15cm}{\caption{\label{fig:4scales}
Scale dependence of the predictions for the first two moments of the photon 
spectrum in RG-improved perturbation theory: variation of $\mu_0$ (solid), 
variation of $\mu_i$ (solid gray), variation of $\mu_h$ (dashed), variation of 
$\mu$ (dashed gray). The light shaded areas show the estimated perturbative
uncertainties.}}}
\end{figure}

In order to obtain a more conservative estimate of the remaining perturbative
uncertainty, one should vary the different scales entering the RG-improved 
expressions independently. This is done in Figure~\ref{fig:4scales}, where we 
explore the sensitivity to variations of the individual matching scales on an 
expanded scale. The underlaid gray bands indicate the total perturbative 
errors we assign. They are obtained by combining the various variations in 
quadrature, ignoring however very low values of the soft scale $\mu_0$, where 
$\alpha_s(\mu_0)$ is so large that perturbation theory deteriorates. Combining 
also the parametric uncertainties in quadrature, and allowing for small 
variations of $m_b$ and $\mu_\pi^2$ about their default values, we obtain for 
$E_0=1.8$\,GeV
\begin{eqnarray}
   \langle E_\gamma\rangle
   &=& (2.287\pm 0.013_{\rm pert}\pm 0.003_{\rm pars})\,\mbox{GeV}
    + 0.44\,\delta m_b + 0.010\,\mbox{GeV}^{-1}\,\delta\mu_\pi^2 \,,
    \nonumber\\
   \sigma_E^2
   &=& (0.0334\pm 0.0051_{\rm pert}\pm 0.0023_{\rm pars})\,\mbox{GeV}^2
    + 0.020\,\mbox{GeV}\,\delta m_b + 0.073\,\delta\mu_\pi^2 \,.
\end{eqnarray}
The central values are in excellent agreement with the results found by the
Belle Collaboration \cite{Koppenburg:2004fz} and collected in 
Table~\ref{tab:data}. This indicates that the values for $m_b$ and $\mu_\pi^2$ 
extracted from $\bar B\to X_c\,l^-\bar\nu$ moment fits are compatible with 
those favored by the $\bar B\to X_s\gamma$ photon spectrum, which by itself is 
a non-trivial test of the heavy-quark expansion. Note that the theoretical 
error estimates are in agreement with the naive estimates presented at the 
end of Section~\ref{sec:sec2intro}. It follows from the first relation that 
the $b$-quark mass can be extracted with exquisitely small theoretical 
uncertainties of only $(\pm 29_{\rm pert}\pm 6_{\rm pars})$\,MeV form the 
average photon energy. From the second moment, the combination 
$\mu_\pi^2+0.27\,\mbox{GeV}\,m_b$ can be extracted with errors of 
$(\pm 0.07_{\rm pert}\pm 0.03_{\rm pars})$\,GeV$^2$. Given the precision 
achieved on $m_b$, these errors essentially determine the precision on the 
extraction of the parameter $\mu_\pi^2$.

\begin{table}
\centerline{\parbox{14cm}{\caption{\label{tab:data}
Experimental results for the first two moments of the $\bar B\to X_s\gamma$ 
photon spectrum, defined with a cut $E_\gamma\ge E_0$. All results refer to 
the $B$-meson rest frame.}}}
\vspace{0.1cm}
\begin{center}
\begin{tabular}{|c|cc|c|}
\hline\hline
$E_0$~[GeV] & $\langle E_\gamma\rangle$~[GeV]
 & $\sigma_E^2$~[$10^{-2}\,\mbox{GeV}^2$] & Reference \\
\hline
1.8 & $2.292\pm 0.027\pm 0.033$ & $3.05\pm 0.79\pm 0.99$
 & Belle \cite{Koppenburg:2004fz} \\
1.9 & $2.321\pm 0.038_{\,-0.038}^{\,+0.017}$\hspace{0.5cm}
 & $2.53\pm 1.01_{\,-0.28}^{\,+0.41}$\hspace{0.4cm}
 & BaBar \cite{Aubert:2005cu} \\
2.0 &  $2.346\pm 0.032\pm 0.011$ & $2.26\pm 0.66\pm 0.20$
 & CLEO \cite{Chen:2001fj} \\
\hline\hline
\end{tabular}
\end{center}
\end{table}

The precision achieved for the mass determination profits greatly from the 
availability of a complete NNLO prediction for the first moment. If we used 
instead only the NLO approximation, the combined perturbative error on 
$\langle E_\gamma\rangle$ would increase from $\pm 13$\,MeV to $\pm 40$\,MeV, 
yielding a theory error of about $\pm 100$\,MeV (from perturbation theory) on 
the extracted value for $m_b$, in agreement with the findings of 
\cite{Neubert:2004dd}. 

The above analysis can be repeated for other values of the cutoff $E_0$, 
however only within certain limits. The default value for the soft scale, 
$\mu_0=\Delta=m_b-2E_0$, is 0.85\,GeV for $E_0=1.9$\,GeV (as used in the BaBar 
analysis in \cite{Aubert:2005cu}), and 0.65\,GeV for $E_0=2.0$\,GeV (as 
employed in the CLEO analysis in \cite{Chen:2001fj}). In the latter case a 
short-distance treatment cannot reasonably be expected to work, and one should 
resort to a description in terms of shape functions, such as 
\cite{Kagan:1998ym,Lange:2005yw}. For $E_0=1.9$\,GeV the applicability of our
approach is marginal, and indeed plots analogous to Figure~\ref{fig:4scales} 
exhibit a more pronounced sensitivity to variations of the low scale $\mu_0$ 
in this case. Accounting for this by a 50\% increase of the perturbative 
error, we find
\begin{eqnarray}
   \langle E_\gamma\rangle
   &=& (2.305\pm 0.020_{\rm pert}\pm 0.003_{\rm pars})\,\mbox{GeV}
    + 0.43\,\delta m_b + 0.016\,\mbox{GeV}^{-1}\,\delta\mu_\pi^2 \,, 
    \nonumber\\
   \sigma_E^2
   &=& (0.0302\pm 0.0077_{\rm pert}\pm 0.0023_{\rm pars})\,\mbox{GeV}^2
    + 0.012\,\mbox{GeV}\,\delta m_b + 0.071\,\delta\mu_\pi^2 \,.
\end{eqnarray}
These theoretical results are in good agreement with the moment measurements 
reported by the BaBar Collaboration \cite{Aubert:2005cu}.

\subsection{Combined moment fits}

\begin{figure}[t]
\begin{center}
\epsfig{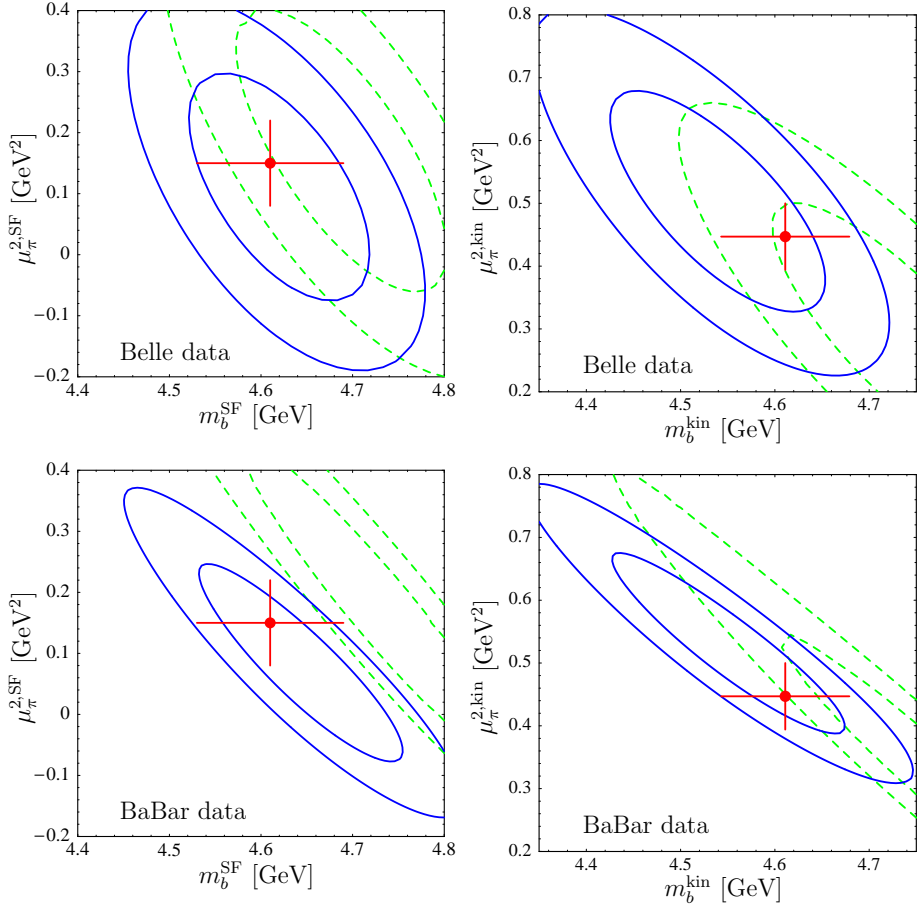}
\end{center}
\vspace{-0.2cm}
\centerline{\parbox{15cm}{\caption{\label{fig:contours}
Fits to the Belle and BaBar data for the moments of the photon spectrum. We 
show contours where $\chi^2=1$ and 2.69, so that projections onto the axes 
yield parameter ranges at 68\% and 90\% confidence level. The fits are 
performed using the shape-function scheme (left) and the kinetic scheme 
(right). The solid (dashed) contour lines refer to the NNLO (NLO) 
approximation. The points with error bars indicate the results obtained from 
the $\bar B\to X_c\,l^-\bar\nu$ moment analysis.}}}
\end{figure}

We are now ready to perform a combined analysis of the experimental data for 
the first two moments of the $\bar B\to X_s\gamma$ photon spectrum with the 
goal to extract the values of the heavy-quark parameters $m_b$ and 
$\mu_\pi^2$. To this end, we define 
\begin{equation}
   \chi^2(m_b,\mu_\pi^2)
   = \sum_{i,j=1,2} (X_i^{\rm exp} - X_i^{\rm th})\,(V^{-1})_{ij}\, 
   (X_j^{\rm exp} - X_j^{\rm th}) \,,
\end{equation}
where $X_1=\langle E_\gamma\rangle$ and $X_2=\sigma_E^2$ are the two 
observables, and $V$ is the covariance matrix containing information about the 
errors and correlations in the measurements of these two quantities
\cite{Koppenburg:2004fz,Aubert:2005cu}. In the theoretical calculation of 
$X_i^{\rm th}$ we keep all theory parameters other than $m_b$ and $\mu_\pi^2$ 
fixed to their default values. Throughout, we use expressions in RG-improved 
perturbation theory. For a given set of measurements, the point where 
$\chi^2=0$ determines the best fit values. Figure~\ref{fig:contours} shows 
contours of $\chi^2=1$ and $\chi^2=2.69$ in the $m_b$--$\mu_\pi^2$ plane 
obtained by fitting the data of the Belle and BaBar Collaborations. We show 
results for the shape-function scheme considered so far, as well as for the 
kinetic scheme (see Appendix~A.3), which has been used, e.g., in the 
$\bar B\to X_c\,l^-\bar\nu$ moment analysis in \cite{Aubert:2004aw}. The solid 
contours refer to the NNLO formulae derived in the present work, while the 
dashed contours correspond to the NLO approximation. The NLO results are 
consistent with those obtained at NNLO when one takes into account theoretical 
uncertainties, which are much larger at NLO. 

Adding the theoretical uncertainties as determined in the previous section 
(errors for the kinetic scheme are determined in an analogous way), we find
\begin{eqnarray}\label{Bellevals}
   m_b^{\rm SF} &=& (4.622\pm 0.099_{\rm exp}\pm 0.030_{\rm th})\,\mbox{GeV}
    \,, \quad\,
   \mu_\pi^{2,{\rm SF}} = (0.108\pm 0.186_{\rm exp}\pm 0.077_{\rm th})\,
    \mbox{GeV}^2 \,, \nonumber\\
   m_b^{\rm kin} &=& (4.543\pm 0.114_{\rm exp}\pm 0.041_{\rm th})\,\mbox{GeV}
    \,, \quad
   \mu_\pi^{2,{\rm kin}} = (0.495\pm 0.176_{\rm exp}\pm 0.085_{\rm th})\,
    \mbox{GeV}^2 \,, \qquad
\end{eqnarray}
from the fit to the Belle data \cite{Koppenburg:2004fz}, and
\begin{eqnarray}\label{BaBarvals}
   m_b^{\rm SF} &=& (4.648\pm 0.111_{\rm exp}\pm 0.047_{\rm th})\,\mbox{GeV}
    \,, \quad\,
   \mu_\pi^{2,{\rm SF}} = (0.076\pm 0.161_{\rm exp}\pm 0.113_{\rm th})\,
    \mbox{GeV}^2 \,, \nonumber\\
   m_b^{\rm kin} &=& (4.556\pm 0.117_{\rm exp}\pm 0.060_{\rm th})\,\mbox{GeV}
    \,, \quad
   \mu_\pi^{2,{\rm kin}} = (0.522\pm 0.143_{\rm exp}\pm 0.122_{\rm th})\,
    \mbox{GeV}^2 \,, \qquad
\end{eqnarray}
from the fit to the BaBar data \cite{Aubert:2005cu}. The fits to the two data 
sets are consistent with each other, but the theoretical errors are smaller 
in the first case due to the lower value of $E_0$ used in the Belle analysis. 
In the shape-function scheme $m_b$ and $\mu_\pi^2$ are defined at 
$\mu_f=1.5$\,GeV, while in the kinetic scheme we adopt the conventional choice 
$\mu_f=1$\,GeV. In all cases there is a strong anti-correlation of the two 
quantities, as can be seen from the figure. 

The values for the heavy-quark parameters determined form the fit to the 
$\bar B\to X_s\gamma$ moments are in excellent agreement with those derived 
from moments in $\bar B\to X_c\,l^-\bar\nu$ decays, which are 
$m_b^{\rm SF}=(4.61\pm 0.08)$\,GeV and 
$\mu_\pi^{2,{\rm SF}}=(0.15\pm 0.07)$\,GeV$^2$ in the shape-function scheme
\cite{Neubert:2004sp}, and $m_b^{\rm kin}=(4.611\pm 0.068)$\,GeV and 
$\mu_\pi^{2,{\rm kin}}=(0.447\pm 0.053)$\,GeV$^2$ in kinetic scheme
\cite{Aubert:2004aw}. These reference values are shown as data points in 
Figure~\ref{fig:contours} for comparison. The combined average values obtained 
from (\ref{Bellevals}) and (\ref{BaBarvals}) are 
$m_b^{\rm SF}=(4.63\pm 0.08)$\,GeV and 
$\mu_\pi^{2,{\rm SF}}=(0.09\pm 0.14)$\,GeV$^2$, and 
$m_b^{\rm kin}=(4.55\pm 0.09)$\,GeV and 
$\mu_\pi^{2,{\rm kin}}=(0.51\pm 0.14)$\,GeV$^2$. However, given that the BaBar 
data are still preliminary and that they employ a higher value of $E_0$, we 
consider the fit to the Belle data as our most reliable result. Combining the 
values in (\ref{Bellevals}) with the ones extracted from the 
$\bar B\to X_c\,l^-\bar\nu$ moment fit yields
\begin{eqnarray}
   m_b^{\rm SF} &=& (4.61\pm 0.06)\,\mbox{GeV} \,, \hspace{0.86cm}
    \mu_\pi^{2,{\rm SF}} = (0.15\pm 0.07)\,\mbox{GeV}^2 \,, \nonumber\\
   m_b^{\rm kin} &=& (4.59\pm 0.06)\,\mbox{GeV} \,, \qquad
    \mu_\pi^{2,{\rm kin}} = (0.45\pm 0.05)\,\mbox{GeV}^2 \,.
\end{eqnarray}

\section{Conclusions}

Moments of the photon energy spectrum in the inclusive radiative decay 
$\bar B\to X_s\gamma$ are sensitive probes of low-scale hadronic dynamics. 
They can be used to extract accurate values for the $b$-quark mass and the 
kinetic-energy parameter $\mu_\pi^2$ of heavy-quark effective theory. Starting 
from an exact QCD factorization formula for the partial $\bar B\to X_s\gamma$ 
decay rate, we have derived improved predictions for the first two moments, 
$\langle E_\gamma\rangle$ and 
$\langle E_\gamma^2\rangle-\langle E_\gamma\rangle^2$, defined with a cut 
$E_\gamma\ge E_0$ on the photon energy. In the region where $\Delta=m_b-2E_0$ 
is large compared with $\Lambda_{\rm QCD}$, a theoretical description without 
recourse to shape (or bias) functions has been achieved. 

The leading terms in the $1/m_b$ expansion of the moments receive 
contributions from the low and intermediate scales $\Delta$ and 
$\sqrt{m_b\Delta}$, but not from the hard scale $m_b$. For these terms, a 
complete scale separation is achieved at next-to-next-to-leading order (NNLO) 
in renormalization-group improved perturbation theory, including two-loop 
matching contributions and three-loop running. Given that 
$\Delta\approx 1$\,GeV is a rather low scale, it is not surprising that the 
NNLO perturbative corrections are numerically significant. They lead to 
significant shifts in the central values of the heavy-quark parameters $m_b$ 
and $\mu_\pi^2$ extracted from a fit to experimental data for the first two 
moments. When the different scales are properly separated using 
renormalization-group techniques, the inclusion of the NNLO corrections helps 
reducing the residual scale uncertainties in the theoretical predictions. This 
allows us to extract the heavy-quark parameters with excellent theoretical 
accuracy, namely $\delta m_b=30$\,MeV and $\delta\mu_\pi^2=0.08$\,GeV$^2$. The 
extracted values are in very good agreement with those derived from moments of 
inclusive $\bar B\to X_c\,l^-\bar\nu$ decay distributions. This agreement is 
gratifying given the different nature of the theoretical framework used to 
analyze these two classes of decays: a conventional operator product expansion 
in the case of $\bar B\to X_c\,l^-\bar\nu$ decay, and QCD factorization in the 
case of $\bar B\to X_s\gamma$.

As the data on the $\bar B\to X_s\gamma$ photon spectrum will become more
accurate in the near future, the tools developed in this work will enable us 
to determine the $b$-quark mass with unprecedented precision. This, in turn, 
will help to reduce the theoretical uncertainties in the determination of 
$|V_{ub}|$ using, e.g., the approach of \cite{Lange:2005yw}.
 
\vspace{0.5cm}\noindent
{\em Acknowledgments:\/}
I am grateful to Thomas Becher, Richard Hill, Bj\"orn Lange, and Gil Paz for 
several useful discussions, to Nikolai Uraltsev for clarifications concerning 
the kinetic scheme, and to Francesca Di Lodovico, Henning Flaecher, and 
Antonio Limosani for communications concerning the BaBar and Belle data. I 
wish to thank the Institute for Advanced Study (Princeton, NJ) for hospitality 
and support as a member for the fall term 2004, during which most of this work 
was completed. This research was supported by the National Science Foundation 
under Grant PHY-0355005, and by the Department of Energy under Grant 
DE-FG02-90ER40542.

\newpage
\section*{Appendices}

\subsubsection*{A.1~~Perturbative expansion of \bm{\eta}}

The expansion of the parameter $\eta$ defined in (\ref{eta}) in RG-improved
perturbation theory can be derived using the perturbative expansions of the 
cusp anomalous dimension and $\beta$ function, which we write as
\begin{equation}
   \Gamma_{\rm cusp}(\alpha_s)
   = \sum_{n=0}^\infty \Gamma_n
    \left( \frac{\alpha_s}{4\pi} \right)^{n+1} , \qquad
   \beta(\alpha_s) = \frac{d\alpha_s}{d\ln\mu}
   = -2\alpha_s \sum_{n=0}^\infty \beta_n
    \left( \frac{\alpha_s}{4\pi} \right)^{n+1} .
\end{equation}
At NNLO, we obtain
\begin{eqnarray}
   \eta &=& \frac{\Gamma_0}{\beta_0}\,\Bigg\{
    \ln\frac{\alpha_s(\mu_0)}{\alpha_s(\mu_i)} 
    + \left( \frac{\Gamma_1}{\Gamma_0} - \frac{\beta_1}{\beta_0} \right)
    \frac{\alpha_s(\mu_0) - \alpha_s(\mu_i)}{4\pi} \nonumber\\
   &&\mbox{}+ \left[ \frac{\Gamma_2}{\Gamma_0} - \frac{\beta_2}{\beta_0}
    - \frac{\beta_1}{\beta_0}
    \left( \frac{\Gamma_1}{\Gamma_0} - \frac{\beta_1}{\beta_0} \right) \right]
    \frac{\alpha_s^2(\mu_0) - \alpha_s^2(\mu_i)}{32\pi^2} + \dots \Bigg\} \,.
\end{eqnarray}
The expansion coefficients of the $\beta$ function to three-loop order in the 
$\overline{\rm MS}$ scheme are \cite{Tarasov:au}
\begin{eqnarray}
   \beta_0 &=& \frac{11}{3}\,C_A - \frac23\,n_f \,, \qquad
    \beta_1 = \frac{34}{3}\,C_A^2 - \frac{10}{3}\,C_A\,n_f - 2 C_F\,n_f \,,
    \nonumber\\
   \beta_2 &=& \frac{2857}{54}\,C_A^3 + \left( C_F^2 - \frac{205}{18}\,C_F C_A
    - \frac{1415}{54}\,C_A^2 \right) n_f
    + \left( \frac{11}{9}\,C_F + \frac{79}{54}\,C_A \right) n_f^2 \,.
\end{eqnarray}
The three-loop expression for the cusp anomalous dimension has recently been 
obtained in \cite{Moch:2004pa}. The expansion coefficients are
\begin{eqnarray}
   \Gamma_0 &=& 4 C_F \,, \qquad
    \Gamma_1 = C_F \left[ \left( \frac{268}{9} - \frac{4\pi^2}{3} \right)
    C_A - \frac{40}{9}\,n_f \right] \,, \nonumber\\
   \Gamma_2 &=& 16 C_F\,\Bigg[ \left( \frac{245}{24} - \frac{67\pi^2}{54}
    + \frac{11\pi^4}{180} + \frac{11}{6}\,\zeta_3 \right) C_A^2
    - \left( \frac{209}{108} - \frac{5\pi^2}{27} + \frac{7}{3}\,\zeta_3
    \right) C_A\,n_f \nonumber\\
   &&\qquad~\mbox{}- \left( \frac{55}{24} - 2\zeta_3 \right) C_F\,n_f
    - \frac{n_f^2}{27} \,\Bigg] \,.
\end{eqnarray}

\subsubsection*{A.2~~Perturbative expansions of the jet and soft functions}

The two-loop matching conditions at the hard-collinear and soft scales are
encoded in the functions $j$ and $s$ defined in (\ref{jdef}) and (\ref{sdef}), 
respectively. At two-loop order, explicit expressions for these quantities
have been given in (\ref{ansatz}). Besides the expansion coefficients of the 
$\beta$ function and cusp anomalous dimension, the results involve the one- 
and two-loop coefficients of anomalous dimensions $\gamma$ and $\gamma^J$, 
which we define as
\begin{equation}
   \gamma^{(J)}(\alpha_s) = \sum_{n=0}^\infty \gamma_n^{(J)}
   \left( \frac{\alpha_s}{4\pi} \right)^{n+1} .
\end{equation}
The two-loop coefficient of the anomalous dimension $\gamma$ entering the 
shape-function evolution kernel has been calculated long ago in 
\cite{Korchemsky:1992xv}, and some errors in this calculation have now been 
corrected \cite{Neubert:2004dd,Gardi:2005yi} (see also the Erratum to 
\cite{Korchemsky:1992xv}). The result is
\begin{equation}\label{littlegamma}
   \gamma_0 = - 2 C_F  \,, \qquad
   \gamma_1 = C_F \left[ \left( \frac{110}{27} + \frac{\pi^2}{18}
    - 18\zeta_3 \right) C_A
    + \left( \frac{4}{27} + \frac{\pi^2}{9} \right) n_f \right] .
\end{equation}
The two-loop anomalous dimension $\gamma^J$ of the jet function in 
soft-collinear effective theory (SCET) has not yet been computed directly. A 
calculation is in progress and has already led to a prediction for the terms 
of order $C_F n_f$ \cite{inprep}. The remaining terms can be deduced by noting 
that the SCET jet function is related to the familiar jet function from 
deep-inelastic scattering. The result is \cite{Neubert:2004dd}
\begin{eqnarray}\label{gammaJ}
   \gamma_0^J &=& - 3 C_F \,,  \nonumber\\
   \gamma_1^J 
   &=& C_F \left[ - \left( \frac32 - 2\pi^2 + 24\zeta_3 \right) C_F
    - \left( \frac{1769}{54} + \frac{11\pi^2}{9} - 40\zeta_3 \right) C_A
    + \left( \frac{121}{27} + \frac{2\pi^2}{9} \right) n_f \right] . \quad
\end{eqnarray}

\subsubsection*{A.3~~Heavy-quark parameters in the kinetic scheme}

The defining relations for the $b$-quark mass and kinetic-energy parameter in 
the kinetic scheme are \cite{Benson:2003kp}
\begin{eqnarray}
   m_b \Big|_{\rm pole}
   &=& m_b(\mu_f) + \frac43\,\mu_f\,\frac{C_F\alpha_s(\mu)}{\pi}
    \left\{ 1 + \frac{\alpha_s(\mu)}{\pi} \left[
    \bigg( \frac12 \ln\frac{\mu}{2\mu_f} + \frac43 \bigg) \beta_0
    + \bigg( \frac{13}{12} - \frac{\pi^2}{6} \bigg) C_A \right] \right\}
    \nonumber\\
   &&\mbox{}+ \frac{\mu_f^2}{2m_b(\mu_f)}\,\frac{C_F\alpha_s(\mu)}{\pi}
    \left\{ 1 + \frac{\alpha_s(\mu)}{\pi} \left[
    \bigg( \frac12 \ln\frac{\mu}{2\mu_f} + \frac{13}{12} \bigg) \beta_0
    + \bigg( \frac{13}{12} - \frac{\pi^2}{6} \bigg) C_A \right] \right\}
    + \dots \,, \nonumber\\
   \mu_\pi^2 \Big|_{\rm pole}
   &=& \mu_\pi^2(\mu_f) - \mu_f^2\,\frac{C_F\alpha_s(\mu)}{\pi}
    \left\{ 1 + \frac{\alpha_s(\mu)}{\pi} \left[
    \left( \frac12 \ln\frac{\mu}{2\mu_f} + \frac{13}{12} \right) \beta_0
    + \left( \frac{13}{12} - \frac{\pi^2}{6} \right) C_A \right] \right\} 
    + \dots \,. \nonumber\\
\end{eqnarray}
The conventional choice for the subtraction scale is $\mu_f=1$\,GeV.

\end{document}